\documentclass[journal, twocolumn]{IEEEtran}

\usepackage[pdftex]{graphicx}
\usepackage[cmex10]{amsmath}
\usepackage{amsfonts}
\usepackage{acronym}
\usepackage{color}
\usepackage{comment}
\usepackage{subcaption}
\usepackage{threeparttable}
\usepackage{booktabs}
\usepackage{amsthm}
\usepackage{upgreek}
\usepackage{multirow}
\usepackage{cite}

\newcommand{\bfb}[1]{\boldsymbol{\rm #1}}
\newcommand{\bfg}[1]{\boldsymbol{#1}}
\newcommand{\nx}{m_x}
\newcommand{\ny}{m_y}
\newcommand{\nxy}{r}
\newcommand{\xs}{\ensuremath{\mbox{$\bfb {x}$}}}
\newcommand{\ys}{\ensuremath{\mbox{$\bfb {y}$}}}

\newcommand{\dirka}{\alpha}
\newcommand{\dirkb}{\beta}
\newcommand{\dirkc}{\gamma}
\newcommand{\tdiE}{\tilde {\bfb E}}
\newcommand{\tdiA}{\tilde {\bfb A}}
\newcommand{\jj}{\jmath}
\newcommand{\zt}{\tilde z}
\newcommand{\T}{^{\scriptscriptstyle \rm T}}

\theoremstyle{definition}
\newtheorem*{theorem*}{Theorem}
\newtheorem{definition}{Definition}[]

\newtheorem{proposition}{Proposition}[]

\acrodef{aiits}[AIITS]{All-Island Irish Transmission System}
\acrodef{bdf}[BDF]{Backward Differentiation Formula}
\acrodef{bdf2}[BDF2]{2-step Backward Differentiation Formula}
\acrodef{bem}[BEM]{Backward Euler Method}
\acrodef{dae}[DAE]{Differential Algebraic Equation}
\acrodef{ddae}[DDAE]{Delay Differential Algebraic Equation}
\acrodef{dde}[DDE]{Delay Differential Equation}
\acrodef{sg}[SG]{Synchronous Generator}
\acrodef{sssa}[SSSA]{Small-Signal Stability Analysis}
\acrodef{fem}[FEM]{Forward Euler Method}
\acrodef{itm}[ITM]{Implicit Trapezoidal Method}
\acrodef{ode}[ODE]{Ordinary Differential Equation}
\acrodef{rk}[RK]{Runge-Kutta}
\acrodef{rk4}[RK4]{Runge-Kutta 4}
\acrodef{tdi}[TDI]{Time Domain Integration}
\acrodef{tds}[TDS]{Time Domain Simulation}
\acrodef{2sdirk}[2S-DIRK]{2-Stage Diagonally Implicit Runge-Kutta}
\acrodef{hh4}[HH4]{Hammer-Hollingsworth~4}
\acrodef{avr}[AVR]{Automatic Voltage Regulator}
\acrodef{tg}[TG]{Turbine Governor}

\title{Small-Signal Stability Analysis of Numerical Integration Methods}

\author{%
  Georgios Tzounas, {\em IEEE Member},
  Ioannis Dassios, and
  Federico Milano, {\em IEEE Fellow}%
  \thanks{The authors are with the School of Electrical and Electronic
    Engineering, University College Dublin, Ireland.  E-mails:
    \{georgios.tzounas, ioannis.dassios, federico.milano\}@ucd.ie.}%
  \thanks{This work is supported by the European Commission, by funding G.~Tzounas and F.~Milano under the project EdgeFLEX, grant
agreement no.~883710; and by Science Foundation Ireland, by funding I.~Dassios and F.~Milano under the Investigator Programme with grant no.~SFI/15/IA/3074.}%
}

\begin{document}

\maketitle \pagestyle{plain} \IEEEpeerreviewmaketitle

\begin{abstract}
  The paper provides a novel framework to study the accuracy and
  stability of numerical integration schemes when employed for the
  time domain simulation of power systems.  A matrix pencil-based
  approach is adopted to evaluate the error between the dynamic modes
  of the power system and the modes of the approximated discrete-time
  system arising from the application of the numerical method.  The
  proposed approach can provide meaningful insights on how different
  methods compare to each other when applied to a power system, while
  being general enough to be systematically utilized for, in
  principle, any numerical method.  The framework is illustrated for a
  handful of well-known explicit and implicit methods, while
  simulation results are presented based on the WSCC $9$-bus system,
  as well as on a $1,479$-bus dynamic model of the All-Island Irish
  Transmission System.
\end{abstract}


\section{Introduction}

\subsection{Motivation}

Time domain simulations are an essential component of power system dynamic analysis and security assessment.  In general, a time domain simulation consists in integrating the dynamic power system model for a set of initial conditions and through a proper numerical method.
The need for the application of a numerical method leads to an approximated representation of the original system's behavior, with the deviation between exact and obtained solution being dependent upon the method's properties and parameters, as well as on the structure of the modeled dynamics.  The goal of this work is to provide a novel and
systematic approach to study the approximation induced to the dynamic modes of power systems by numerical \acf{tdi} methods.

\subsection{Literature Review}

A dynamic power system model is conventionally described by a set of
non-linear and stiff \acfp{dae} \cite{stott:1979}.  
The \ac{tdi} of a
power system often relies on the implementation of an implicit numerical method, since explicit schemes -- such as the \ac{fem} --
perform poorly for stiff problems.  
The implicit method most commonly utilized in power system dynamic simulations is arguably the \ac{itm}, yet a number of schemes have been proposed to achieve the best compromise between accuracy and efficiency of simulation, see \cite{1972dommel, stott:1979, 1989marti, 1994astic, 1995paserba,
  2009noda, stiffdecay, hh4:2013}.  For example, some studies propose to combine or substitute the \ac{itm} with a hyperstable scheme, such as the \ac{bem}, with a scope to improve the handling of discontinuities and avoid undamped numerical oscillations under large time steps \cite{1989marti, stiffdecay}.

The precision of a \ac{tdi} is typically assessed on the basis of certain metrics, such as the local and global truncation errors of the numerical scheme employed.  Truncation errors provide a good measure
of the deviation between exact and numerically computed trajectories and are the standard criterion used for the implementation of automatic step size and order control techniques \cite{1994astic,
  1995paserba}.  However, the ability of a \ac{tdi} method to prevent the exponential growth of truncation errors cannot be predicted with
the truncation errors.  Information on the latter is given instead from the characterization of \ac{tdi} methods according to their properties of numerical stability.  
As a matter of fact, the major
advantage of implicit over explicit methods is that they outperform in terms of numerical stability.  For example, the \ac{itm} is symmetrically A-stable, i.e.~it converges for stable and diverges for
unstable trajectories, whereas the \ac{fem} is unstable, i.e.~it always diverges for sufficiently large time steps.

The classical approach to stability characterization of a \ac{tdi} method is to check its response when applied to a linear test equation.  Consequently, the main limitation of this approach is that it is only qualitative, since it does not involve the dynamics of the specific model to be integrated, and it is thus not suitable for accuracy assessment.  On the contrary, this work is concerned with the problem of providing a unified framework, based on \acf{sssa}, to study the accuracy and stability of numerical methods applied for the \ac{tdi} of power systems.  This problem is tackled by introducing a generic model of numerical method that admits as special cases the most important families of methods, the behavior of which is then
analyzed by studying the associated matrix pencils
\cite{book:eigenvalue}.  The proposed formulation allows quantifying the spurious distortion that a \ac{tdi} method introduces to the dynamic modes of the power system model to which it is applied, as well as systematically realizing relevant analysis tools already
available in the literature.  In this vein we cite
\cite{2002borodulin}, which presents a tool to assess the numerical approximation of the motion of simple linear networks by means of distortion maps.

\subsection{Contributions}

The specific contributions of the paper are as follows.

\begin{itemize}
\item A general numerical stability analysis framework based on matrix
  pencils that, in principle, is applicable to any numerical \ac{tdi}
  scheme.
\item The proposed framework is utilized to evaluate the numerical
  distortion introduced by \ac{tdi} methods to the dynamic modes of
  power system models.
  % The paper demonstrates that the proposed
  % approach can be applied, in principle, to any numerical method.
\item For certain \ac{rk} methods, it is also shown that the pencil to
  be studied emerges as an extension of the method's growth function.
\item A discussion on how the proposed approach can be employed to
  estimate useful upper time step bounds that satisfy certain accuracy
  criteria, as well as to provide fair computational-burden
  comparisons of different methods.
\end{itemize}

It is important to note that one cannot ``compare'' the proposed framework to a specific integration method.  Rather, one can use the
proposed approach to define the numerical stability properties of such
an integration method.

\subsection{Organization}

The remainder of the paper is organized as follows.
Section~\ref{sec:tdi} describes the dynamic power system model and
provides preliminaries on the numerical \ac{tdi}.  The proposed
framework to study the stability and accuracy of power system \ac{tdi}
is presented in Section~\ref{sec:sssa}.  The case studies are
discussed in Section~\ref{sec:casestudies} based on the well-known
WSCC system and a dynamic model of the \ac{aiits}.  Finally,
conclusions are drawn in Section~\ref{sec:conclusion}.

\section{Integration of Power System Model}
\label{sec:tdi}

\subsection{Power System Model}

The mathematical model that describes the dynamics of a power system can be formulated as follows:
\begin{equation}
  \begin{aligned}
    \label{eq:singular}
    \bfb E \, {\xs}'(t) &= \bfg \phi( \xs(t) ) \, , 
  \end{aligned}    
\end{equation}
where $\bfb E \in \mathbb{R}^{\nxy \times \nxy}$;
$\xs : [0,\infty) \rightarrow \mathbb{R}^{\nxy}$ is the column vector
of the system's variables and ${\xs}'$ denotes the time derivative of
${\xs}$; $\bfb \phi : \mathbb{R}^{\nxy} \rightarrow \mathbb{R}^{\nxy}$
is a set of non-linear functions that defines the equations of the
system.
Discrete variables in \eqref{eq:singular} are modeled implicitly,
i.e., each discontinuous change in the system leads to a jump from
\eqref{eq:singular} to a new continuous set of equations in the same
form.
A relevant special case is when \eqref{eq:singular} is formulated as a
set of explicit \acp{dae}, i.e.:
% respectively;
% 
\begin{equation}
  \bfb E = 
  \begin{bmatrix}
    \bfg I_{\nx} & \bfg 0_{\nx,\ny} \\
    \bfg 0_{\ny,\nx} & \bfg 0_{\ny,\ny} \\
  \end{bmatrix}
  , \,
  \bfg \phi( \xs(t) ) = 
  \begin{bmatrix}
    \bfg f( \bfg  x(t) , \bfg y(t)) \\
    \bfg g( \bfg  x(t) , \bfg y(t)  )  \\
  \end{bmatrix}  \, ,
  \label{eq:dae:explicit:mat}
\end{equation}
and $\xs(t) = [\bfg x\T(t) \ \bfg y\T(t)]\T$, where
$\bfg x : [0,\infty) \rightarrow \mathbb{R}^{\nx}$ and
$\bfg y: [0,\infty) \rightarrow \mathbb{R}^{\ny}$ are the state and
algebraic variables; and
$\bfg f : \mathbb{R}^{\nx+\ny} \rightarrow \mathbb{R}^{\nx}$,
$\bfg g : \mathbb{R}^{\nx+\ny} \rightarrow \mathbb{R}^{\ny}$, are
non-linear functions that define the differential and algebraic equations, respectively; $\bfg I_{\nx}$ denotes the $\nx \times \nx$
identity matrix and $\bfg 0_{\nx,\ny}$ the $\nx \times \ny$ zero
matrix.  Equivalently, one has:
\begin{equation}
  \begin{aligned}
    \label{eq:dae:explicit}
    {\bfg x}'(t)  &= \bfg f( \bfg  x(t) , \bfg y (t) ) \, , \\
    \bfg 0_{\ny,1} &=  \bfg  g( \bfg  x(t) , 
    \bfg y (t) ) \, .
  \end{aligned}    
\end{equation}
Formulation \eqref{eq:dae:explicit} is the standard model employed in
the literature for transient and voltage stability studies
\cite{kundur:94}.  Yet, the main results of this work hold also for the
more general model \eqref{eq:singular} and, thus, \eqref{eq:singular}
is the starting point considered in this paper.

\subsection{Numerical Integration}

A \ac{tdi} method for power systems is a discrete-time approximation
employed to solve system \eqref{eq:singular} for a defined time period
and set of initial conditions.  We propose the following generic model
to describe a \ac{tdi} method.

\vspace{1mm} {\definition { In an implicit form, a \ac{tdi} method
    applied to system \eqref{eq:singular} can be described by a %set of
    %non-linear difference equations
    discrete-time system, as follows:
    \begin{equation}
      \label{eq:tdi:implicit}
      \bfg 0_{\nxy,1} = \bfg \eta(\xs_{t}, \xs_{t-h}, \xs_{t-a_1h},\xs_{t-a_2h},\ldots,\xs_{t-a_\rho h})
      \, ,
    \end{equation}
    where $h$ is the integration time step size, which can be constant
    or varying; $\nxy=\nx+\ny$;
    $\bfg \eta : \mathbb{R}^{(\rho+2)\nxy} \rightarrow
    \mathbb{R}^{\nxy}$ is a vector of non-linear functions; $\xs_{t} : \mathbb{N}^*h\rightarrow\mathbb{R}^{\nxy}$;
    and
    $a_i>0$, $a_i\neq1$, $i=1,2,\ldots,\rho$.  }
  \label{definition:method}
  \vspace{1.5mm}
}

The discrete-time system \eqref{eq:tdi:implicit} covers most elements
of the two largest and most important families of \ac{tdi} methods,
namely \ac{rk} and linear multistep methods.  We clarify here that
\eqref{eq:tdi:implicit} is expressed in an implicit form but this
should not be confused with the method being implicit or not.  In
fact, both explicit and implicit methods can be represented in the
form of \eqref{eq:tdi:implicit}.  With this regard, we propose the
following definition of explicit numerical methods applied to system
\eqref{eq:singular}.

\vspace{1mm}
{\definition
  {
    If, without resorting to further approximations, system \eqref{eq:tdi:implicit} can be equivalently rewritten in the form:
    \begin{equation}
      \label{eq:map:explicit}
      \bfb E \, \xs_{t} = \bfg \theta(\xs_{t-h}, \xs_{t-a_1h},\xs_{t-a_2h},\ldots,\xs_{t-a_\rho h})
      \, ,
    \end{equation}
    where $\bfb E$ is given by \eqref{eq:dae:explicit:mat}; and $\bfg \theta : \mathbb{R}^{(\rho+2)\nxy} \rightarrow \mathbb{R}^{\nxy}$; then it describes an explicit numerical method.  Otherwise, the method is implicit.
  }
  \label{definition:explicit}
  \vspace{1.5mm}
}

Implicit methods involve an extra computation compared to explicit methods, i.e.~they require the solution of system \eqref{eq:tdi:implicit} for $\xs_{t}$.  
This computation is done iteratively at every step of the integration. For instance, the $i$-th iteration of Newton's method when applied to \eqref{eq:tdi:implicit} is:
\begin{equation}
  \label{eq:map:newton}
  \xs^{(i)}_{t} = \xs^{(i-1)}_{t}
  - 
  \left [ \frac{\partial \bfg \eta^{(i-1)}}{\partial \xs^{(i-1)}_{t}}
  \right ]^{-1}
  \bfg \eta^{(i-1)}
  \, ,
\end{equation}
% % % 
where ${\partial \bfg \eta}/{\partial \xs_{t}}$ denotes the Jacobian matrix of \eqref{eq:tdi:implicit}.  
The fact that such computational step is not required by explicit methods is the reason why the latter are still the option preferred by some software tools.  We cite, for example, the use of the explicit modified Euler method in \cite{psse_pag2}.  Hence, for completeness, this paper discusses both explicit and implicit methods. 

\subsection{Problem Stiffness and Small-Signal Model}

The \ac{tdi} of a power system constitutes a stiff problem, i.e., the time constants that define the differential equations of the model span multiple time scales.  A 
measure of stiffness is given by the \textit{stiffness ratio} of the corresponding small-signal model.
Let $\xs_o$ be an equilibrium point of \eqref{eq:singular}.  Then, linearization around $\xs_o$ gives:
\begin{equation}
  \label{eq:dae:lin}
  \bfb E \, \Delta \xs'(t) = 
  \bfb A \, \Delta \xs(t) \, ,
\end{equation}
where $\bfb A = \partial \bfg \phi/\partial \xs$ and $\Delta \xs(t) = \xs-\xs_o$.  The eigenvalues of \eqref{eq:dae:lin} are the solutions of the characteristic equation:
\begin{equation}
  \label{eq:dae:chareq}
  {\rm det}(s \bfb E - \bfb A) = 0 \, ,
\end{equation}
where the family of matrices $s \bfb E - \bfb A$ parameterized by
$s \in \mathbb{C}$ is called the \textit{matrix pencil} of system
\eqref{eq:dae:lin} \cite{book:eigenvalue}.  In total, the pencil
$s \bfb E - \bfb A$ has $\upnu = {\rm rank}(s \bfb E - \bfb A)$ finite
eigenvalues plus the infinite eigenvalue with multiplicity
$\nxy - \upnu$.  Moreover, \eqref{eq:dae:lin} is
\textit{asymptotically stable} if and only if the real parts of all
finite eigenvalues $s^*$ of $s \bfb E - \bfb A$ satisfy
${\rm Re}(s^*) < 0$.  In practice, the eigenvalues of
$s \bfb E - \bfb A$ are computed numerically, see \cite{app10217592}.
We finally provide the following definition.
{\definition{ Assume that \eqref{eq:dae:lin} is
    asymptotically stable and let $s_i=\alpha_i+\jj\beta_i$,
    $i=1,2,\ldots,\upnu$, be the $i$-th finite eigenvalue of
    $s \bfb E-\bfb A$.  Let also $\sigma_{\max} = \max\{|\alpha_i|\}$,
    $\sigma_{\min} = \min\{|\alpha_i|\}$ denote the maximum, minimum
    exponential decay rates of the system, respectively.  Then, the
    stiffness ratio of \eqref{eq:dae:lin} is
    \cite{lambert1991numerical}:\footnote{Spurious zero eigenvalues
      due to the arbitrariness of the reference angle and the
      redundancy of one or more machine rotor angle equations (see the
      discussion in \cite{7390066}) are not taken into account in
      Definition~\ref{definition:stiffness}.   }
\begin{equation}
  \label{eq:dae:stiffness}
  \mathcal{S} = \frac{\sigma_{\max}}{\sigma_{\min}} \, .
\end{equation}
}
\label{definition:stiffness}
}

It is relevant to note that the definition of stiffness is not unique. For example, an alternative definition may also take into account the effect of the imaginary parts $\beta_i$, e.g.~by defining as measure of stiffness the ratio of the finite eigenvalues with largest and smallest magnitude. 

Apart from its presence in \eqref{eq:dae:stiffness}, the maximum
exponential decay rate $\sigma_{\max}$ is an index commonly employed
by software tools to make an heuristic estimation of the maximum
admissible integration time step based on empirical rules, to prevent
either that the fastest dynamics of the system are filtered out, or,
in the case of an explicit method, that convergence is compromised
\cite{1992demello}.  On the contrary, in this paper we systematically
evaluate the error of the \ac{tdi} method in approximating the dynamic
modes of the system which allows extracting upper time step bounds
with higher accuracy.

\subsection{Classical Stability Analysis}

This section briefly recalls the classical approach to stability
analysis of numerical \ac{tdi} methods.  The stability of a numerical
method for ordinary differential equations is traditionally tested and
classified by applying the method to Dahlquist's test equation:
\begin{equation}
  \label{eq:test}
  \xi'(t)=\lambda \; \xi(t) \, ,
\end{equation}
where
$\lambda \in \mathbb{C}$.
% $\lambda<0$. 
% Equation \eqref{eq:test}
% is often referred to in the literature as 
Let apply an integration method 
to \eqref{eq:test} so that:
\begin{equation}
  \label{eq:stabfunc}
  \xi_t=\mathcal{R}(\lambda h) \; \xi_{t-h} \, ,
\end{equation}
where $\mathcal{R}(\lambda h)$ is the method's \textit{growth} or
\textit{stability function}.  Then, the stability region of the method
is defined by the set:
\begin{equation}
  \label{eq:test:region}
  \{ \lambda \in \mathbb{C} \, :
  \  |\mathcal{R}(\lambda h)|<1 \}
  \, .
\end{equation}
As an example, 
consider the 
application of the \ac{itm} to \eqref{eq:test}:
\begin{equation}
  \label{eq:itm:test}
  \xi_t=\xi_{t-h} + 0.5h\lambda \xi_{t-h} + 0.5h\lambda \xi_{t}
  \, ,
\end{equation}
which can be equivalently written in the form of \eqref{eq:stabfunc}, where:
\begin{equation}
  \label{eq:itm:stability}
  \mathcal{R}(\lambda h) =
  \frac{1+0.5\lambda h}{1-0.5\lambda h} \, . 
\end{equation}
From \eqref{eq:test:region}, \eqref{eq:itm:stability}, we have that the stability region of the \ac{itm} is the left half of the $S$-plane.

For \ac{rk} methods, the growth function can be written as \cite{1979scherer}:
\begin{equation}
 \label{eq:stabfunc:but}
 \mathcal{R}(\lambda h) =
 \frac{{\rm det}(\bfg I_\rho - \lambda h\bfb Q
+ \lambda h
\bfb e_\rho
\bfb r)}{{\rm det}(\bfg I_\rho - \lambda h\bfb Q)} \, ,
\end{equation}
where $\rho$ is the method's number of
stages; $\bfb e_\rho$ is the 
$\rho \times 1$ vector of ones; and
$[\bfb Q\T \  \bfb r\T ]\T$,
is the method's generating matrix,
$\bfb Q \in \mathbb{R}^{\rho\times \rho}$,
$\bfb r \in \mathbb{R}^{1\times \rho}$.
% \[
%   \begin{array}{l|l}
      %       \bfb c & \bfb Q \\
      %       \midrule
      %                      & \bfb b\T \\
      %     \end{array} \, .
      %       \]
Note that explicit \ac{rk} methods
have 
      %       a lower triangular matrix $\bfb Q$ so that 
${\rm det}(\bfg I_\rho - h \lambda \bfb Q)=1$, 
and thus their growth function 
is a polynomial of $\lambda h$.
On the other hand, the growth function of
an %$\rho$-stage 
implicit \ac{rk} method is a quotient of two polynomials of $\lambda h$.	

\section{Matrix Pencil-based Numerical Analysis}
\label{sec:sssa}

\subsection{Proposed Approach}
\label{sec:proposed}

In this section we provide a general approach to study the numerical
distortion caused by \ac{tdi} methods to the dynamic modes of system \eqref{eq:singular}.  First, we prove that the approximation introduced by any numerical method applied to a system in the form of
\eqref{eq:singular} can be studied through a linear matrix pencil. Consider the discrete-time system \eqref{eq:tdi:implicit} and assume
for simplicity but without loss of generality that $h$ is constant. Then, linearization of the system around the equilibrium $\xs_o$ of
\eqref{eq:singular}, which is also a fixed point of
\eqref{eq:tdi:implicit}, gives:
\begin{align}
  \hspace{-1mm}
  \bfg 0_{\nxy,1} =&
 \frac{\partial\bfg \eta}{\partial\xs_{t}} 
 \Delta \xs_{t}
 +
 \frac{\partial\bfg \eta}{\partial\xs_{t-h}} \Delta \xs_{t-h}+
 \frac{\partial\bfg \eta}{\partial\xs_{t-a_1h}} \Delta \xs_{t-a_1h}
 \nonumber
\\
&
 +\frac{\partial\bfg \eta}{\partial\xs_{t-a_2h}}
 \Delta \xs_{t-a_2h}
 +
 \ldots+
 \frac{\partial\bfg \eta}{\partial\xs_{t-a_\rho h}}
 \Delta\xs_{t-a_\rho h} \, .
 \label{eq:tdi:implicit:lin}
\end{align}

We provide the following proposition.

\vspace{1mm} {\proposition { The stability properties of system
\eqref{eq:tdi:implicit:lin} can be assessed by studying the stability of a linear 
discrete-time 
system %of difference equations 
in the form:
    \begin{equation}
      \label{eq:tdi:sssa}
      \tdiE \, \ys_{t} = 
      \tdiA \, \ys_{t-h}
      \, .
    \end{equation}
  }
  \label{proposition:tdi:pencil}
  % \vspace{0.5mm}
}

The proof of Proposition~\ref{proposition:tdi:pencil} is provided in the Appendix.  Then, the stability of \eqref{eq:tdi:sssa}
can be seen through the eigenvalues of the matrix pencil
$\zt \tdiE - \tdiA$.  In particular, \eqref{eq:tdi:sssa} is
asymptotically stable if and only if all finite eigenvalues $\zt^*$ of its pencil $\zt \tdiE - \tdiA$ lie within the open unit disc, or equivalently, $|\zt^*| < 1 $.

The eigenvalues of $\zt \, \tdiE - \tdiA$ represent, in the $Z$-plane, the small-disturbance dynamic modes of \eqref{eq:singular} as
approximated by the numerical method \eqref{eq:tdi:implicit}. Let $\zt_k$ be an eigenvalue of $\zt \, \tdiE - \tdiA$ approximating the
$k$-th dynamic mode of the power system model, which is represented by
the finite eigenvalue $s_k=\alpha + \jj \beta$ of $s \bfb E - \bfb A$.
Then, the two eigenvalues become directly comparable by mapping the
one to the domain of the other.  Mapping $\zt_k$ from the $Z$-plane to
the $S$-plane, we get:
\begin{equation}
  \label{eq:map:ztos}
  \tilde s_k = \frac{1}{h} {\rm log} (\zt_k)  
  = \tilde\alpha + \jj \tilde\beta
  \, ,
\end{equation}
where ${\rm log}(\cdot)$ denotes the complex logarithm.  Then, the numerical distortion caused to the $k$-th mode by the \ac{tdi} method is:
\begin{align}
  \label{eq:error:s}
  {d}_{s,k} = \tilde s_k - s_k \, .
\end{align}
The distortion caused to the damping of the $k$-th mode is:
\begin{align}
  \label{eq:error:zeta}
  {d}_{\zeta,k} = 
  \tilde \zeta_k - \zeta_k \,  ,
\end{align}
where $\zeta_k=-{\alpha}/({\alpha^2+\beta^2})$.  Positive (negative) values of ${d}_{\zeta,k}$ indicate that the mode is overdamped (underdamped).

\subsection{Illustrative Examples}
\label{sec:methods}

This section discusses the matrix pencils that characterize the
stability and accuracy of some well-known integration \ac{tdi}
methods.  In particular, six methods are considered, namely (i)
\ac{fem}, (ii) \ac{rk4}, (iii) \ac{bem}, (iv) \ac{itm}, (v)
\ac{2sdirk} and (vi) \ac{bdf2}.  These methods are also employed for the case studies of Section~\ref{sec:casestudies}.

\subsubsection*{\acf{fem}}

The \ac{fem} is the simplest among all integration schemes.  When applied to system \eqref{eq:singular}, the \ac{fem} reads:
\begin{equation}
  \label{eq:fem}
  \bfb E \xs_{t}= 
  \bfb E \xs_{t-h} 
  + h \bfg \phi ( \xs_{t-h}) 
  \, ,
\end{equation}
where $\xs'(t)$ is approximated with the finite difference formula
$(\xs_t-\xs_{t-h})/h$.  Linearization of \eqref{eq:fem} around $\xs_o$
gives:
\begin{equation}
  \label{eq:fem:lin}
  \bfb E \Delta\xs_{t}= 
  \bfb E \Delta\xs_{t-h} 
  + 
  h \bfb A \Delta\xs_{t-h}
  \, .
\end{equation}
Equivalently, \eqref{eq:fem:lin} can be rewritten as a discrete-time system in the form of \eqref{eq:tdi:sssa} with pencil $\zt \tdiE - \tdiA$, where $\ys_{t}\equiv \Delta\xs_{t}$ and:
\begin{equation}
  \begin{aligned}
    \label{eq:fem:pencil}
    \tdiE &= {\bfb E} \, ,
    \quad
    \tdiA = {\bfb E} + h {\bfb A} \, .    
  \end{aligned}
\end{equation}

\subsubsection*{\acf{rk4}}

The classical \ac{rk4} is a fourth-order method and is the most well-known explicit \ac{rk} method.  Applied to \eqref{eq:singular},
the \ac{rk4} method reads:
\begin{align}
  \label{eq:rk4}
  \hspace{-4mm}
  \bfb E \, \xs_{t} &= \bfb E \, \xs_{t-h}
  +\frac{h}{6}(\bfg k_1+2 \bfg k_2+2\bfg k_3+\bfg k_4)  \, ,\\
  \bfg k_1 &= \bfg \phi(\xs_{t-h}) \, , \hspace{1.75cm}
 \bfg k_2 = \bfg \phi(\xs_{t-h}+0.5h\bfg k_1 ) \, , \nonumber\\
  \bfg k_3 &= \bfg \phi(\xs_{t-h}+0.5h \bfg k_2 ) \, , \quad
 \bfg k_4 = \bfg \phi(\xs_{t-h}+h \bfg k_3 ) 
 \nonumber \, . 
\end{align}
Linearization of \eqref{eq:rk4}	yields the following expressions:
\begin{equation}
 \begin{aligned}\label{eq:rk4:sssa}
\bfg k_1 &= \bfb A  \Delta\xs_{t-h} \, , 
\hspace{2cm}
\bfg k_2= \bfb A (\Delta\xs_{t-h}+0.5h \bfg k_1) \, , \\
\bfg k_3&= \bfb A ( \Delta\xs_{t-h}+0.5h \bfg k_2)  \, ,
\quad
\bfg k_4=\bfb A ( \Delta\xs_{t-h}+h \bfg k_3)  \, . \\
 \end{aligned}
\end{equation}
Equivalently, the linearized method can be written in the form of \eqref{eq:tdi:sssa} with matrix pencil $\zt \tdiE - \tdiA$, where
$\ys_{t}\equiv \Delta\xs_{t}$ and:
\begin{equation}
  \begin{aligned}
    \label{eq:rk4:pencil}
    \tdiE &= {\bfb E}  \, , \\
    \tdiA &= {\bfb E} + h{\bfb A}+\frac{(h{\bfb A})^2}{2} + \frac{(h{\bfb A})^3}{6} + \frac{(h{\bfb A})^4}{24}  \, .
  \end{aligned}
\end{equation}

\subsubsection*{\acf{bem}}

The \ac{bem} is the implicit variant of the \ac{fem} and is a hyperstable method with stability region the part of the $S$-plane that is outside the unit disk centered at 1.  When applied to system
\eqref{eq:singular}, the \ac{bem} reads:
\begin{equation}
  \label{eq:bem}
  \bfb E \xs_{t} = 
  \bfb E \xs_{t-h} 
  + h \bfg \phi (\xs_{t}) \, .
\end{equation}
Linearization of \eqref{eq:bem} leads to a discrete-time system in the
form of \eqref{eq:tdi:sssa}, where $\ys_{t}\equiv \Delta\xs_{t}$ and:
\begin{equation}
  \begin{aligned}
    \label{eq:bem:pencil}
    \tdiE &= {\bfb E}- h {\bfb A}  \, , 
    \quad
    \tdiA = {\bfb E}   \, .
  \end{aligned}
\end{equation}

\subsubsection*{\acf{itm}}

The \ac{itm} can be interpreted as the weighted sum of the \ac{fem} and \ac{bem} with equal weights for the two methods.  Applied to system \eqref{eq:singular}, the \ac{itm} reads:
\begin{equation}
  \label{eq:itm}
  \bfb E \xs_{t} = 
  \bfb E \xs_{t-h} 
  + 0.5h \bfg \phi (\xs_{t-h}) 
  + 0.5h \bfg \phi (\xs_{t})
  \, .
\end{equation}
Linearization of \eqref{eq:itm} 
leads to a system in the form of 
\eqref{eq:tdi:sssa}, where
$\ys_{t}\equiv \Delta\xs_{t}$ and:
\begin{equation}
  \begin{aligned}
    \label{eq:itm:pencil}
    \tdiE &= {\bfb E}- 0.5h {\bfb A} \, ,
    \quad
    \tdiA = {\bfb E}+ 0.5h {\bfb A} \, .
  \end{aligned} 
\end{equation}

Note that permitting for unequal weights in \eqref{eq:itm} leads to a
generalized version of the \ac{itm} commonly referred to as the Theta
method \cite{1995paserba}.  As a byproduct of the adopted pencil-based approach, we can show that the \ac{fem}, \ac{bem}, \ac{itm}, as well
as all elements of the Theta method belong to the wider family of methods whose pencils arise from the application of a linear spectral
transform to $s \bfb E - \bfb A$.  Most importantly, studying such generalized family of pencils allows revealing the elements possessing
certain qualitative properties, such as a certain class of numerical stability.  As an example, in this paper we obtain conditions under which an element of the family is symmetrically A-stable.  The
relevant propositions and their proofs are provided in the Appendix.

\subsubsection*{\acf{2sdirk}}

Diagonally implicit \ac{rk} methods is a family of methods suitable for the solution of stiff initial value problems.  In this paper, we
consider the \ac{2sdirk} method proposed in \cite{2009noda} for the simulation of electromagnetic transients.  The method reads:
\begin{equation}
  \begin{aligned}
    \label{eq:2sdirk}
    \bfb E\, \xs_{t+(\dirka-1) h} &= \,\bfb E \xs_{t-h} + \dirka h \,
    \bfg \phi(\xs_{t+(\dirka-1) h}) \, , \\
    \bfb u_{t-h} &= \dirkb \xs_{t-h} + \dirkc \, \xs_{t+(\dirka-1) h} \, , \\
    \bfb E \, \xs_{t} &= \bfb E \, \bfb u_{t-h}
    +\dirka h \, \bfg \phi(\xs_{t})
    \, ,\\
  \end{aligned}
\end{equation}
with $\dirka=1-1/\sqrt{2}$, $\dirkb=-\sqrt{2}$, $\dirkc=1+\sqrt{2}$.
Linearizing \eqref{eq:2sdirk}:
\begin{equation}
  \begin{aligned}
    \label{eq:2sdirk:sssa}
    \bfb E\Delta\xs_{t+(\dirka-1) h} &= \bfb E \Delta\xs_{t-h} + \dirka h
    \bfb A \Delta\xs_{t+(\dirka-1) h} \, , \\
    \Delta\bfb u_{t-h} &= \dirkb \Delta\xs_{t-h} + \dirkc \, 
    \Delta\bfb x_{t+(\dirka-1) h} \, , \\
    \bfb E \Delta\xs_{t} &= \bfb E \Delta\bfb u_{t-h}
    +\dirka h \bfb A \Delta\xs_{t}  \, .\\
  \end{aligned}
\end{equation}
By eliminating $\Delta\bfb u_{t-h}$, one can rewrite \eqref{eq:2sdirk:sssa} as follows:
\begin{equation}
  \begin{aligned}
    \label{eq:2sdirk:sssa2}
    \bfb E\xs_{t+(\dirka-1) h} &= \bfb E \Delta\xs_{t-h} + \dirka h \,
    \bfb A \Delta\xs_{t+(\dirka-1) h} \, , \\
    \bfb E  \Delta\xs_{t} &= \dirkb \bfb E \Delta\xs_{t-h} + \dirkc \bfb E 
    \Delta\bfb x_{t+(\dirka-1) h}
    +\dirka h  \bfb A \Delta\xs_{t}  \, ,\\
  \end{aligned}
\end{equation}
or equivalently:
\begin{align}
\label{eq:2sdirk:sssa4_1}
% \bfg 0_{\nxy,1}
(\bfb E-\dirka h \bfb A) &\Delta\xs_{t+(\dirka-1)h}
 = \bfb E \Delta\xs_{t-h} \, , \\
\label{eq:2sdirk:sssa4_2}
(\bfb E -\dirka h \bfb A )& \Delta\xs_{t} = ( \bfb E-\dirka \dirkb h \bfb A )  \Delta\xs_{t+(\dirka-1) h} \, ,
\end{align}
where we have replaced $\dirkb+\dirkc=1$.  Substituting
\eqref{eq:2sdirk:sssa4_1} to \eqref{eq:2sdirk:sssa4_2} leads to a
system in the form of \eqref{eq:tdi:sssa}, where $\ys_{t}\equiv \Delta\xs_{t}$ and:
\begin{equation}
  \begin{aligned}
    \label{eq:2sdirk:sssa5}
    \tdiE &= 
    \bfb E - \dirka h \bfb A  \, , \\
    \tdiA &= 
    ( \bfb E-\dirka \dirkb h \bfb A )
    (\bfb E-\dirka h \bfb A)^{-1} \bfb E
    \,  .
  \end{aligned}
\end{equation}

\subsubsection*{\acf{bdf2}}

The backward differentiation formulas is a family of implicit, linear multistep methods.  In this paper, we consider the \ac{bdf2} which, when applied to \eqref{eq:singular}, reads:
\begin{equation}
  \label{eq:bdf2}
  \bfg 0_{\nxy,1} = 
  \bfb E \xs_{t} -\frac{4}{3} \bfb E \xs_{t-h} + \frac{1}{3} \bfb E \xs_{t-2h}
  - \frac{2}{3} h \bfg \phi (\xs_{t}) 
  % - d \bfg \phi ( \xs_t) 
  \, .
\end{equation}
Linearization of \eqref{eq:bdf2} gives:
\begin{equation}
  \label{eq:bdf2:sssa2}
  (\bfb E - \frac{2}{3} h \bfb A) \Delta\xs_{t} = 
  \frac{4}{3} \bfb E \Delta\xs_{t-h} - \frac{1}{3} \bfb E \Delta\xs_{t-2h}
  % - d \bfg \phi ( \xs_t) 
  \, .
\end{equation}
Adopting the notation:
\begin{equation}
  \bfb {y}_{t} = 
  \begin{bmatrix}
    \Delta\xs_{t-h} \\
    \Delta\xs_{t}  \\
  \end{bmatrix}  , \
  \bfb {y}_{t-h} =
  \begin{bmatrix}
    \Delta\xs_{t-2h} \\
    \Delta\xs_{t-h}  \\
  \end{bmatrix} 
  , \, 
  \nonumber
\end{equation}
the system can be written in the form of \eqref{eq:tdi:sssa}, where:
\begin{equation}
  \tdiE=
  \begin{bmatrix}
    \bfg I_\nxy &  \bfg 0_{\nxy,\nxy} \\
    \bfg 0_{\nxy,\nxy} & \bfb E  -\frac{2}{3}h \bfb A\\
  \end{bmatrix}   , \
  \tdiA =
  \begin{bmatrix}
    \bfg 0_{\nxy,\nxy} & \bfg I_\nxy \\
    - \frac{1}{3} \bfb E & \frac{4}{3} \bfb E \\
  \end{bmatrix} 
  .
  \nonumber
\end{equation}

\subsection{Link to Growth Function}
\label{sec:link}

In this section, we discuss the link between the growth function of a \ac{rk} method and the corresponding matrix pencil $\zt \tdiE - \tdiA$
that arises if the method is applied for the \ac{tdi} of
\eqref{eq:singular}.
First, consider the test equation \eqref{eq:test} and write $\lambda$ as a ratio of two values, i.e., $\lambda=\mu_1/\mu_2$.  Then,
\eqref{eq:stabfunc:but} can be rewritten as the ratio of two functions
of $\mu_1$, $\mu_2$ and $h$, as follows:
\begin{equation}
  \label{eq:stabfunc:mu1mu2}
  \mathcal{R}(\lambda h) :=
  \mathcal{F}(\mu_1,\mu_2,h)=
  \frac{\mathcal{N}(\mu_1,\mu_2,h)}{\mathcal{D}(\mu_1,\mu_2,h)}
  \, ,
\end{equation}
where 
\begin{equation}
  \begin{aligned}
    \label{eq:stabfunc:ND}
    {\mathcal{N}}(\mu_1, \mu_2, h) &=
    {\rm det}(\mu_2 \bfg I_\rho - h \mu_1 \bfb Q
    + h \mu_1 \bfb e_\rho \bfb b\T)
    \, , \\
    {\mathcal{D}}(\mu_1, \mu_2, h) &=
    {\rm det}(\mu_2 \bfg I_\rho - h \mu_1 \bfb Q)
    \, .
  \end{aligned}
\end{equation}
Using \eqref{eq:stabfunc:mu1mu2}, \eqref{eq:stabfunc} becomes:
\begin{equation}
  \label{eq:stabfunc:gen}
  \mathcal{D}(\mu_1,\mu_2,h) x_t=\mathcal{N}(\mu_1,\mu_2,h) x_{t-h} 
  \, ,
\end{equation}
and hence, the numerical stability of the method can be equivalently seen through the pencil
$\zt \mathcal{D}(\mu_1,\mu_2,h) - \mathcal{N}(\mu_1,\mu_2,h)$.

For explicit \ac{rk} methods, as well as for certain 
implicit methods, including the \ac{bem} and \ac{itm},
the discussion above can be extended for system \eqref{eq:dae:lin} integrated through \eqref{eq:tdi:sssa}.  Observing that matrices $\tdiA$, $\tdiE$, are functions of $\bfb A$, $\bfb E$ and $h$, and
extending the scalar functions $\mathcal{N}$, $\mathcal{D}$ to the corresponding matrix functions $\bfg {\mathcal{N}}$,
$\bfg {\mathcal{D}}$, we find that:
\begin{equation}
  \begin{aligned}
    \label{eq:stabfunc:AE}
    \tdiE &= 
    \bfg {\mathcal{D}}(\bfb A, \bfb E, h)
    \, , 
    \quad
    \tdiA = 
    \bfg {\mathcal{N}}(\bfb A, \bfb E, h)
    \, .
  \end{aligned}
\end{equation}

As a consequence of \eqref{eq:stabfunc:AE}, the pencils associated with the \ac{rk} methods can be readily obtained by extending known results about their growth functions.  For example, setting $\lambda=\mu_1/\mu_2$ in \eqref{eq:itm:test}, one obtains that the growth function of the \ac{itm} is given by \eqref{eq:stabfunc:mu1mu2}, where:
\begin{equation}
  \begin{aligned}
    \label{eq:stabfunc:itm1}
    {\mathcal{D}(\mu_1,\mu_2,h)} &=
    {\mu_2-0.5\mu_1 h} \, , \\
    \mathcal{N}(\mu_1,\mu_2,h) &=
    {\mu_2+0.5\mu_1 h} 
    \, , \nonumber 
  \end{aligned}
\end{equation}
and hence, consistently with Section~\ref{sec:methods}, one obtains that:
\begin{equation}
  \begin{aligned}
    \label{eq:stabfunc:AE:itm}
    \tdiE &= 
    \bfg {\mathcal{D}}(\bfb A, \bfb E, h) = 
    \bfb E -0.5h\bfb A
    \, , \nonumber \\
    \tdiA &= 
    \bfg {\mathcal{N}}(\bfb A, \bfb E, h)
    = \bfb E +0.5h\bfb A
    \, .
  \end{aligned}
\end{equation}

\subsection{Validity of \ac{sssa}}
\label{sec:validity}

The paper relies upon the linearization of systems \eqref{eq:singular}
and \eqref{eq:tdi:implicit} at a steady state solution $\xs_o$.
Strictly speaking, thus, the proposed approach is valid only around
$\xs_o$.  With this regard, the following remarks are relevant.

In the neighborhood of $\xs_o$, \eqref{eq:error:s} and
\eqref{eq:error:zeta} provide precise measures of the modes' numerical
approximation given a time step or, the other way around, provide the
required step size to achieve a certain accuracy.  A method that does
not fulfill the user's requirements in view of these measures can be
discarded without the need for further calculations.  Thus, the proposed
tool can be very useful when comparing between different methods or
testing potential new numerical schemes on their suitability for
\ac{tdi} of a given power system model.  Last but not least, the
proposed tool requires only the calculation of the associated matrix
pencils and thus it allows testing methods whose full implementation
in the time domain routine may be an involved procedure.

The structure of the dynamic modes and the stiffness of a power system
model are features that do not change dramatically by varying the
operating point, and thus we stress that the proposed measures are
also rough yet accurate estimates of the amount of approximation
introduced by \ac{tdi} methods under varying operating conditions,
also owing to the qualitative properties of the methods which remain
unchanged, such as their class of numerical stability.  Therefore, the
analysis does not need to be repeated often.  Other works that have
faced a similar problem yet for different application are
e.g.~\cite{arriaga:82_2, book:chow:13, tzounas:onestepdelay:2021}.
  
\begin{figure*}[ht!]
  \begin{subfigure}{.333\linewidth}
    \centering
    \resizebox{\linewidth}{!}{\includegraphics{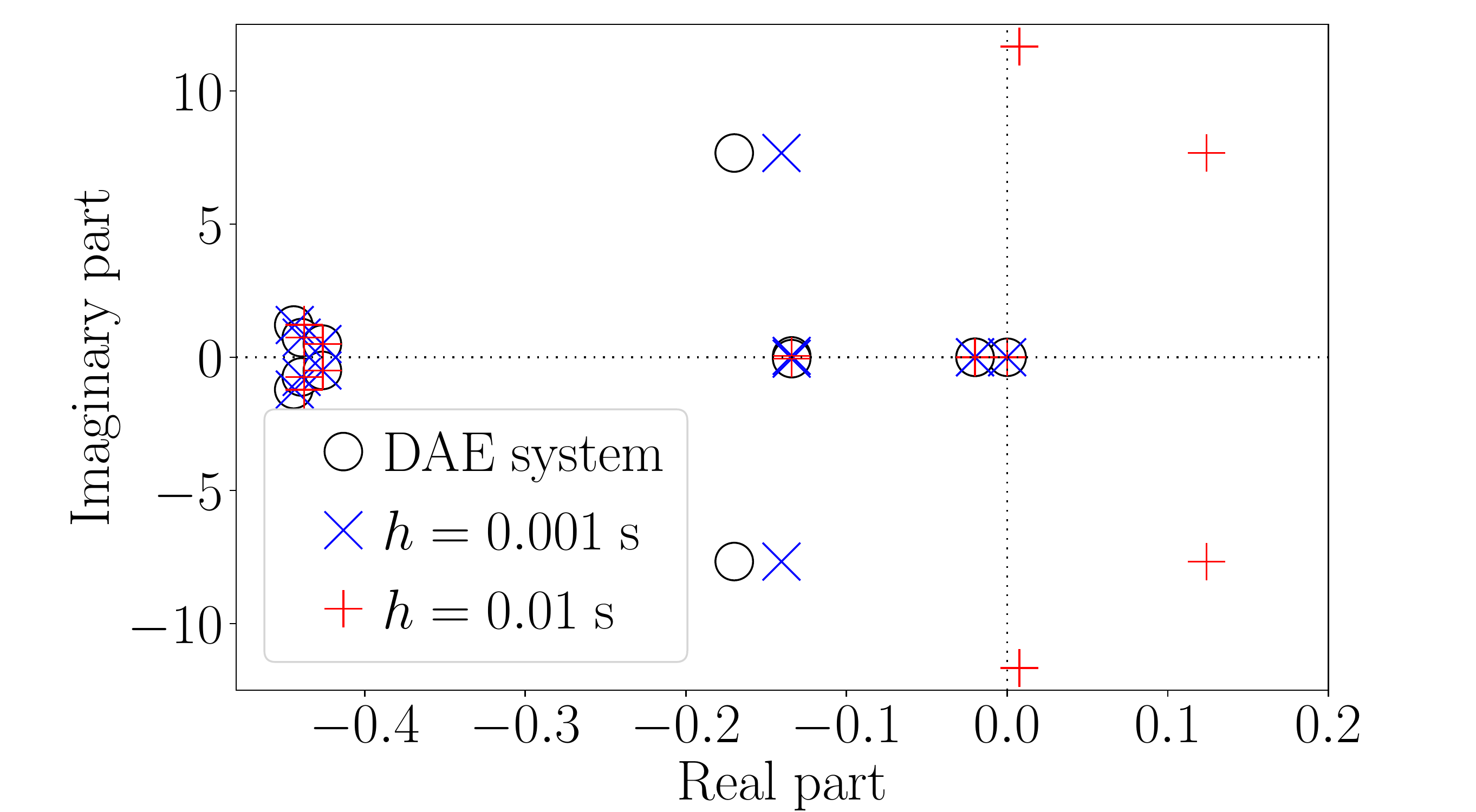}}
    \caption{FEM.}
    \label{fig:dae:fem}
  \end{subfigure}
  \begin{subfigure}{.333\linewidth}
    \centering
    \resizebox{\linewidth}{!}{\includegraphics{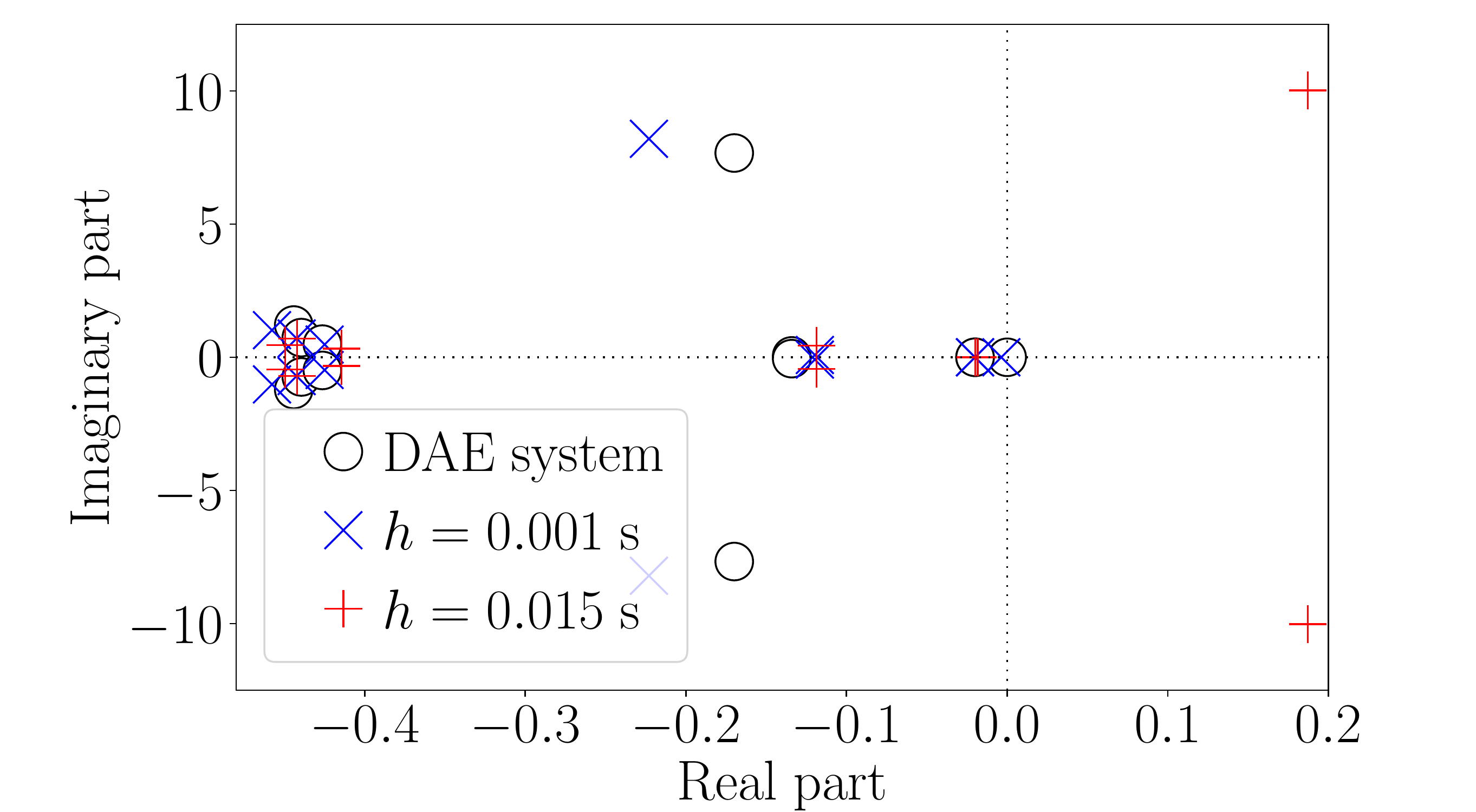}}
    \caption{RK4.}
    \label{fig:dae:rk4}
  \end{subfigure}
  \begin{subfigure}{.333\linewidth}
    \centering
    \resizebox{\linewidth}{!}{\includegraphics{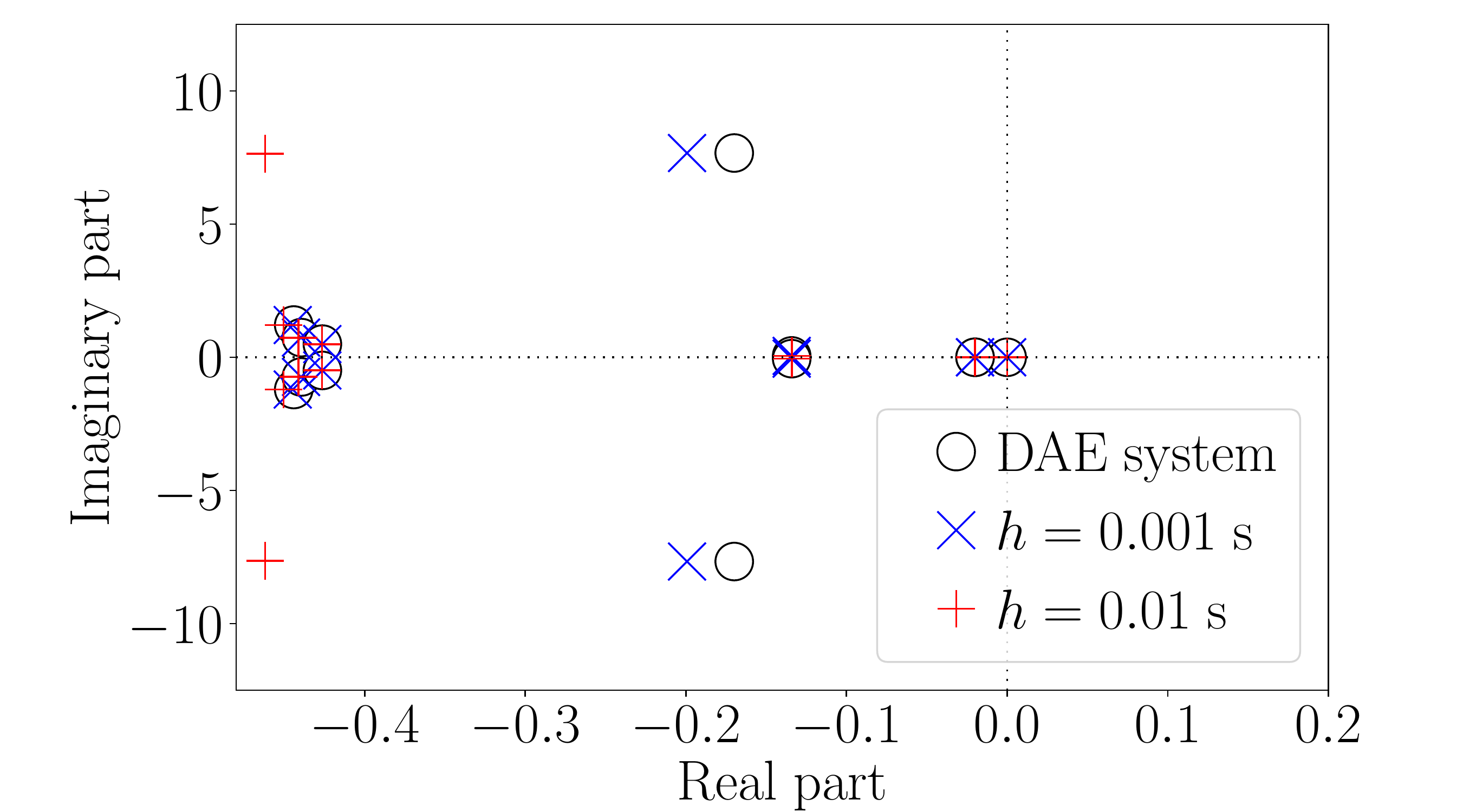}}
    \caption{BEM.}
    \label{fig:dae:bem}
  \end{subfigure}
  \begin{subfigure}{.333\linewidth}
    \centering
    \resizebox{\linewidth}{!}{\includegraphics{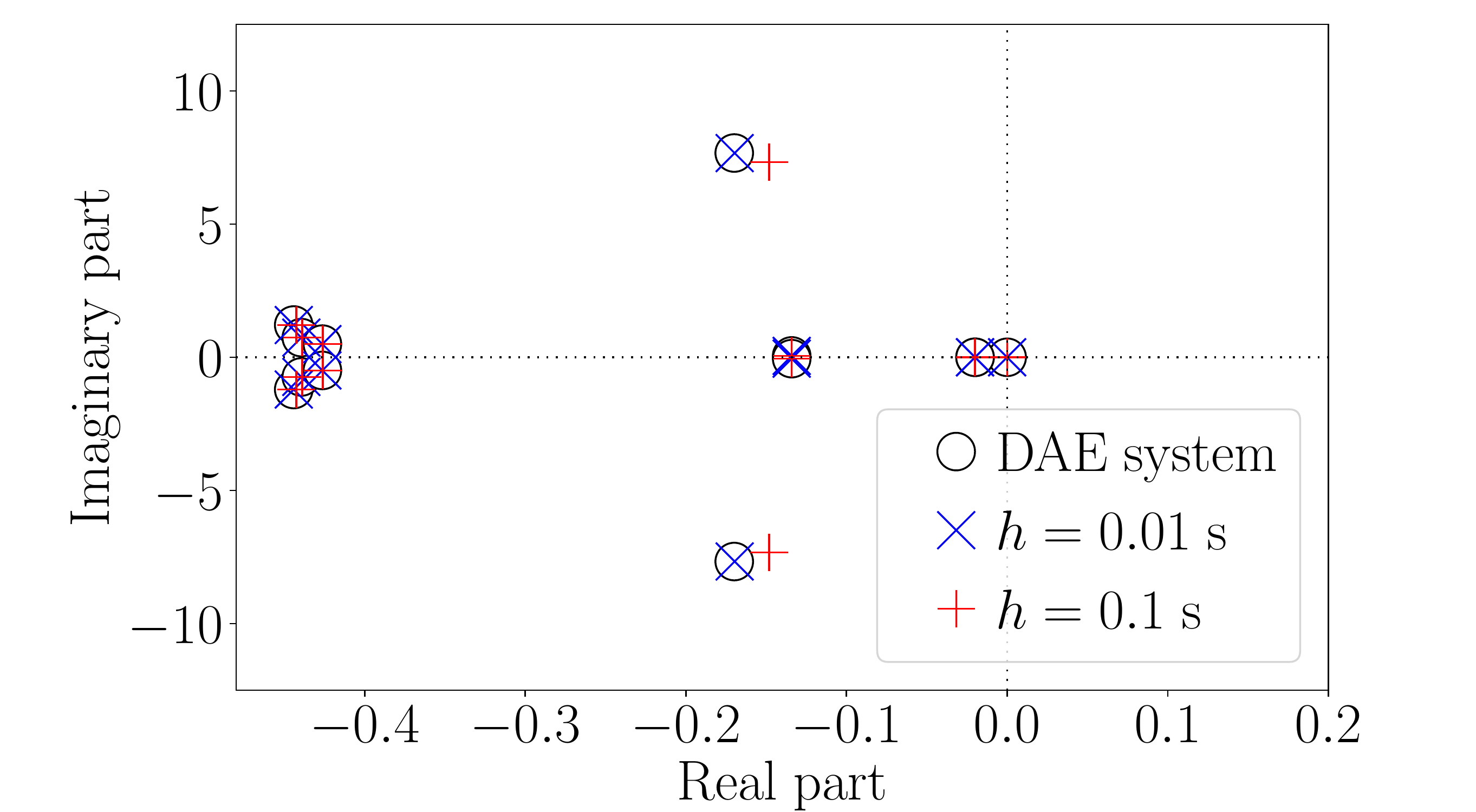}}
    \caption{ITM.}
    \label{fig:dae:itm}
  \end{subfigure}
  \begin{subfigure}{.333\linewidth}
    \centering
    \resizebox{\linewidth}{!}{\includegraphics{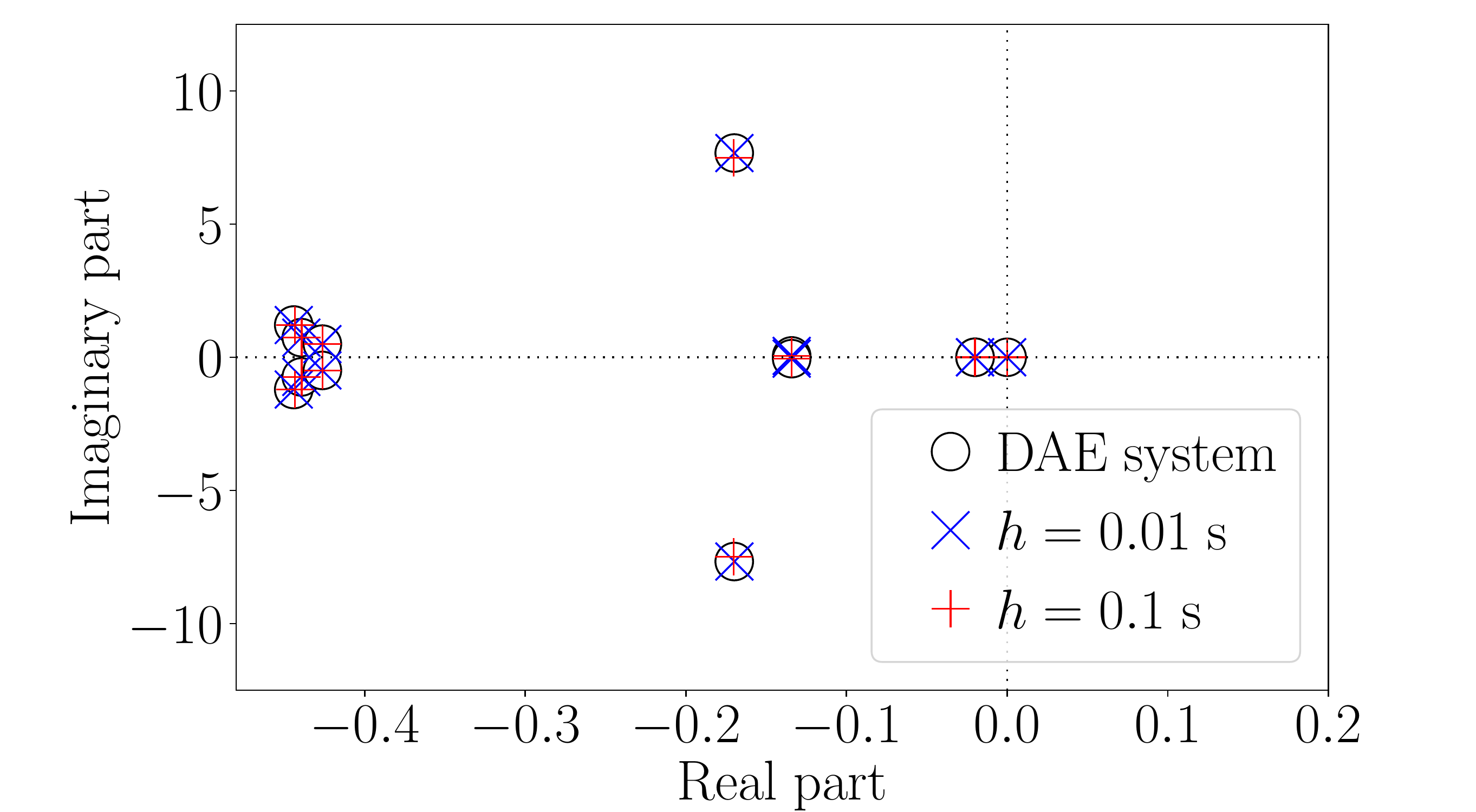}}
    \caption{2S-DIRK.}
    \label{fig:dae:2sdirk}
  \end{subfigure}
  \begin{subfigure}{.333\linewidth}
    \centering
    \resizebox{\linewidth}{!}{\includegraphics{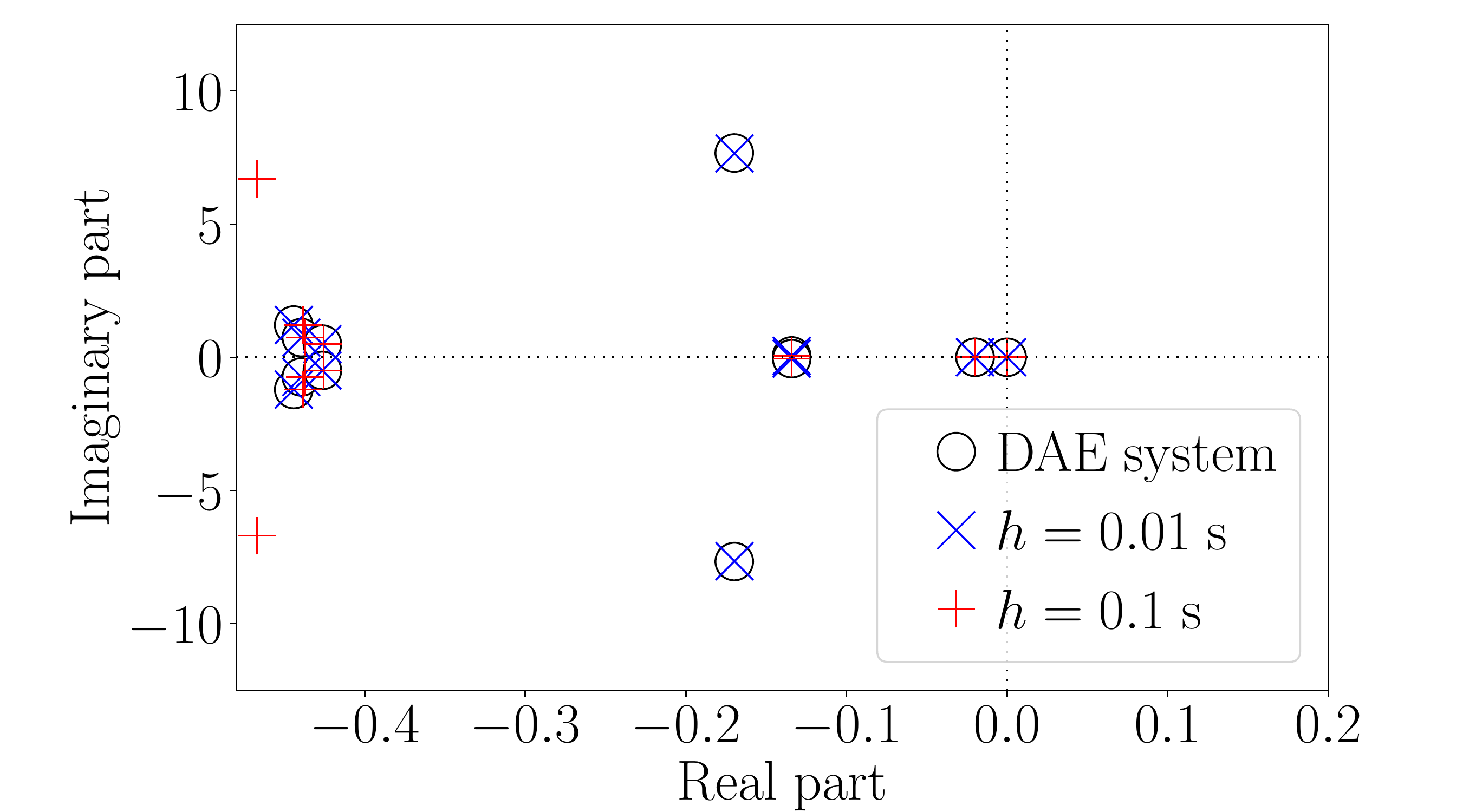}}
    \caption{BDF2.}
    \label{fig:dae:bdf}
  \end{subfigure}
  \caption{WSCC system: Eigenvalue analysis of numerical \ac{tdi} methods.}
  \label{fig:wscc:eig}
  \vspace{-3mm}
\end{figure*}

\section{Case Studies}
\label{sec:casestudies}

The simulation results provided in this section illustrate important features of the proposed framework to study the stability and accuracy
of numerical methods applied for the \ac{tdi} of power systems.  The case study in Section~\ref{sec:wscc} is based on the well-known WSCC $9$-bus system \cite{sauer:1998}, whereas Section~\ref{sec:aiits} considers a realistic model of the \ac{aiits}.

Simulations are carried out using Dome \cite{vancouver}.  The version
of Dome employed in this paper depends on ATLAS~3.10.3 for dense vector/matrix operations; CVXOPT~1.2.5 for sparse matrix operations;
and KLU~1.3.9 for sparse matrix factorizations.  The eigenvalues of matrix pencils are calculated using LAPACK \cite{lapack}.
All simulations are executed on a 64-bit Linux operating system running on 2 quad-core Intel Xeon 3.5 GHz CPUs, and 12 GB of RAM.

\subsection{WSCC 9-Bus System}
\label{sec:wscc}

This section presents simulation results based on the WSCC $9$-bus system.  The system comprises 6 transmission lines and 3 medium voltage/high voltage transformers; 3~\acp{sg} represented by
fourth-order, two-axis models and equipped with \acp{tg} and \acp{avr}. In transient conditions, loads are modeled as constant admittances.  In total, the system's \ac{dae} model includes $39$
state and $57$ algebraic variables.

The small-disturbance dynamics of the system are represented by the eigenvalues of the pencil $s \bfb E - \bfb A$.  Eigenvalue analysis
shows that the system is stable when subjected to small disturbances, with the fastest and slowest dynamics represented by the real
eigenvalues $-1000$ and $-0.02$, respectively, which gives a stiffness
ratio $\mathcal{S} = 5 \cdot 10^4$.

\begin{figure}[ht!]
  \begin{subfigure}{1\linewidth}
    \centering
    \resizebox{\linewidth}{!}{\includegraphics{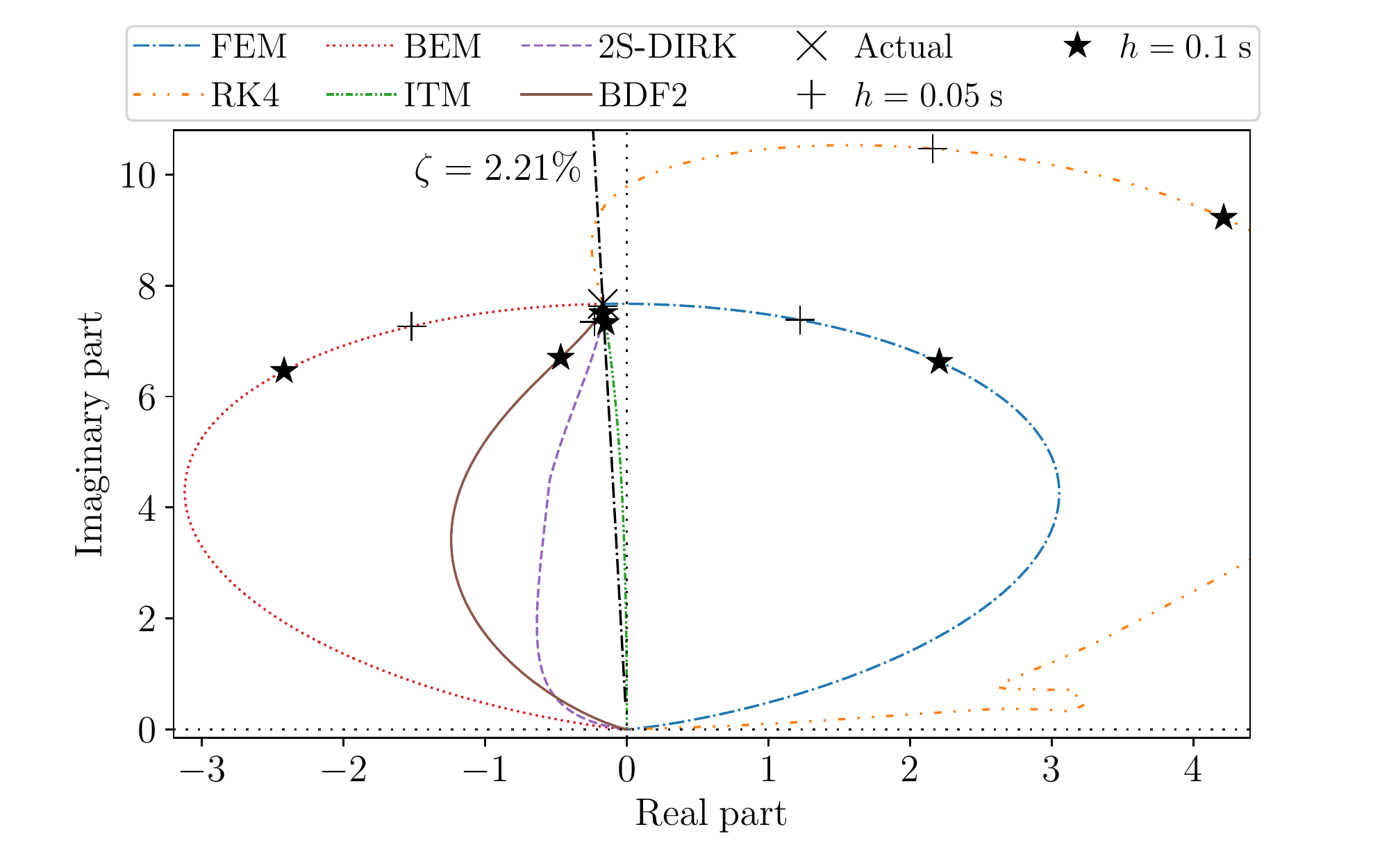}}
    \caption{Root locus.}
    \label{fig:wscc:locus}
  \end{subfigure}
  \begin{subfigure}{1\linewidth}
    \centering
    \resizebox{\linewidth}{!}{\includegraphics{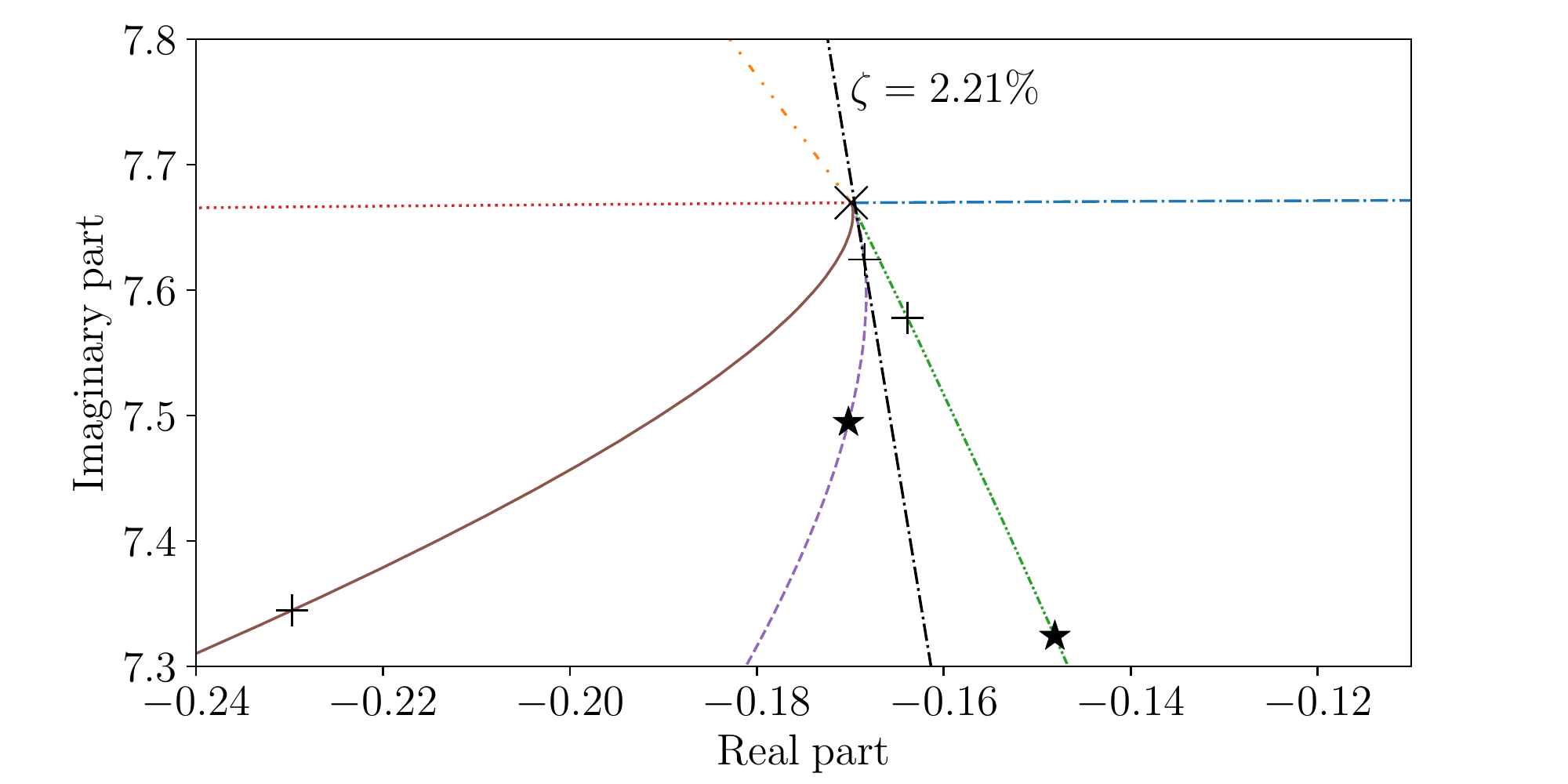}}
    \caption{Root locus (close-up).}
    \label{fig:wscc:locus:zoom}
  \end{subfigure}
  \caption{WSCC system dominant mode: Root locus of numerical
    approximation as the time step is increased.}
  \label{fig:wscc:locus12}
  \vspace{-3mm}
\end{figure}

We consider the six numerical methods discussed in
Section~\ref{sec:methods}.  For each method, we calculate the associated pencil $\zt \, \tdiE - \tdiA$ and its eigenvalues, which
are then mapped to the $S$-plane according to \eqref{eq:map:ztos} and
compared to the eigenvalues of $s \bfb E - \bfb A$.  The comparison
results for different time step sizes are presented in Fig.~\ref{fig:wscc:eig}.  As expected, for a sufficiently large $h$ the explicit methods are destabilized.  Varying $h$ allows determining
the maximum time step before numerical stability is lost. In particular, the time step margin of the \ac{fem} and the \ac{rk4} for
the WSCC system are obtained as $0.005$~s and $0.011$~s, respectively.
For larger time steps, there is at least one dynamic mode for which
$\alpha<0$ and $\tilde\alpha>0$ in \eqref{eq:error:s} and thus any \ac{tdi} executed with such step values is guaranteed to diverge.  The
figure also shows that both the \ac{bem} and the \ac{bdf2} overdamp the dynamics of the system, which is again as expected.  In addition,
varying $h$ allows estimating the upper time step bound for which the overdamping is less than a certain prescribed degree.  For instance,
if it is required that the overdamping of all dynamic modes of the system is less than $d_\zeta=1$\% (see also
eq.~\eqref{eq:error:zeta}), then the upper bounds of $h$ for the
\ac{bem} and \ac{bdf2} are $0.002$~s and $0.051$~s, respectively.
Finally, among all methods considered, the \ac{2sdirk} shows the
highest accuracy, while the \ac{itm} also shows very good accuracy for time steps smaller than $10^{-1}$~s.
We focus on the dominant dynamic mode of the system, i.e.~the local electromechanical oscillation of the \ac{sg} connected to bus~2.  In
the eigenvalue analysis, this mode is represented by the complex pair
$-0.1699 \pm \jj 7.6696$ with damping ratio $2.21$\%. The root loci in
Fig.~\ref{fig:wscc:locus12} illustrate the accuracy of the \ac{tdi} methods in approximating the mode as $h$ increases.  The figure shows the route of explicit methods towards instability as well as of hyperstable methods to overdamped regions.  From the close-up shown in
Fig.~\ref{fig:wscc:locus:zoom}, it is seen that, interestingly, the \ac{rk4} introduces a slight overdamping for small enough step sizes. Comparing the \ac{itm} with the \ac{2sdirk}, which are both symmetrically A-stable, we see that under the same step the \ac{2sdirk} is more accurate. In addition, for steps smaller than $0.05$~s, the \ac{2sdirk} follows precisely the mode's damping, whereas for larger steps, it introduces a slight overdamping. On the other hand, the \ac{itm} underdamps the mode, which for a large enough $h$ leads to sustained numerical oscillations.  For the sake of example, Table~\ref{tab:wscc:zetah} gives the damping distortion $d_\zeta$ introduced by all methods for $h=0.05$~s.  Finally, as $h$ increases, the distorted eigenvalue approaches zero for all methods.  This is consistent with \eqref{eq:map:ztos}, whereby
substituting the limit case $h \rightarrow \infty$, one has $\tilde s_k \rightarrow 0$.

\begin{table}[ht!]
  \centering
  \setlength{\tabcolsep}{5.2pt}
  \renewcommand{\arraystretch}{1.1}
  \caption{WSCC system dominant mode: Damping distortion for
    $h=0.05$~s; and time step leading to $|d_s|=0.1$.}
  \label{tab:wscc:zetah}
  \begin{threeparttable}
    \begin{tabular}{cccccccc}  
      \toprule[0.1pt]%\toprule[0.1pt]
      & \ac{fem} & \ac{rk4} & \ac{bem} & \ac{itm} & \ac{2sdirk} & \ac{bdf2} \\          
      \midrule[0.1pt]
      $d_\zeta$~[\%]
      & \multirow{2}{6mm}{-18.5}
      & \multirow{2}{6mm}{-22.4}
      & \multirow{2}{6mm}{18.2}
      & \multirow{2}{8mm}{-0.052}
      & \multirow{2}{8mm}{-0.005}
      & \multirow{2}{5mm}{0.9}  \\
      ($h=0.05$~s) \\
      \midrule
      %	$d_\zeta$~[\%] 
      %	& -18.5 & -22.4 
      %	& 18.2 & -0.052 & -0.005 & 0.9 \\
      $h$~[s]
      & \multirow{2}{6mm}{0.003}
      & \multirow{2}{6mm}{0.0002}
      & \multirow{2}{6mm}{0.003}
      & \multirow{2}{6mm}{0.052} 
      & \multirow{2}{6mm}{0.075} 
      & \multirow{2}{6mm}{0.026} 
      \\
      ($|d_s|=0.1$) \\
      % $h$~[s] & 0.003 & 0.0002 & 0.003 & 
      % 0.052 & 0.075  & 0.026 \\			
      \bottomrule[0.1pt]%\bottomrule[0.1pt]
    \end{tabular}
  \end{threeparttable}
  \vspace{-3mm}
\end{table}	

\begin{figure}[ht!]
  \centering
  \resizebox{\linewidth}{!}{\includegraphics{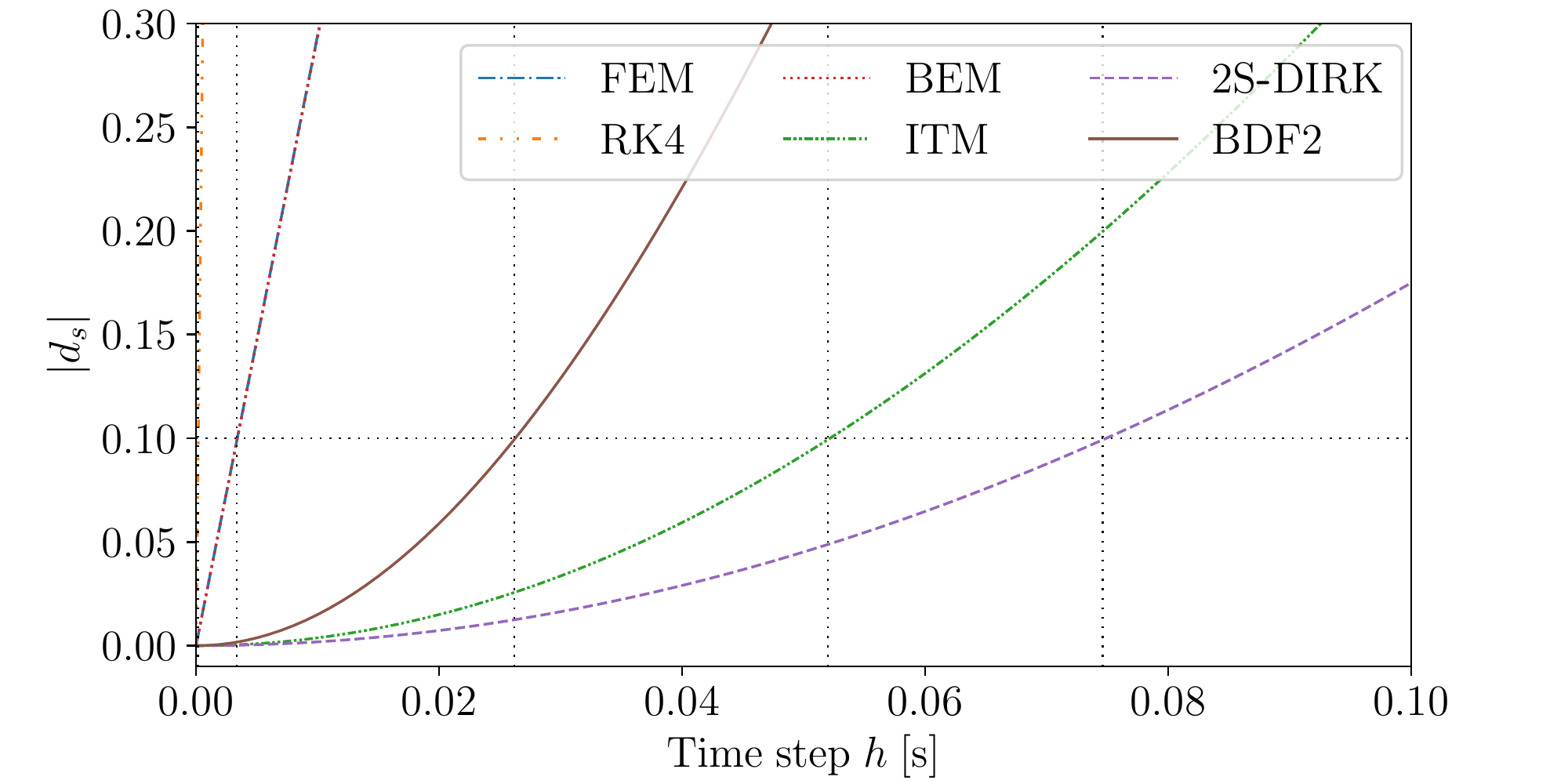}}
  \caption{Dominant mode of WSCC system: Magnitude of numerical distortion.}
  \label{fig:wscc:error}
\end{figure}

For the same mode, the magnitude of numerical distortion as a function
of $h$ is depicted in Fig.~\ref{fig:wscc:error}.  The \ac{bem} and the
\ac{fem}, being the former the implicit version of the latter, cause
practically the same amount of distortion for every $h$, yet in
opposite directions, with the \ac{bem} leading to overdamping and the
\ac{fem} to instability (see Fig.~\ref{fig:wscc:locus}).  Figure
\ref{fig:wscc:error} allows determining the time step that introduces a certain amount of numerical distortion to the mode.  For example,
the time step of each method leading to $|d_s|=0.1$ is given in Table~\ref{tab:wscc:zetah}.

\begin{figure}[ht!]
  \centering
  \resizebox{\linewidth}{!}{\includegraphics{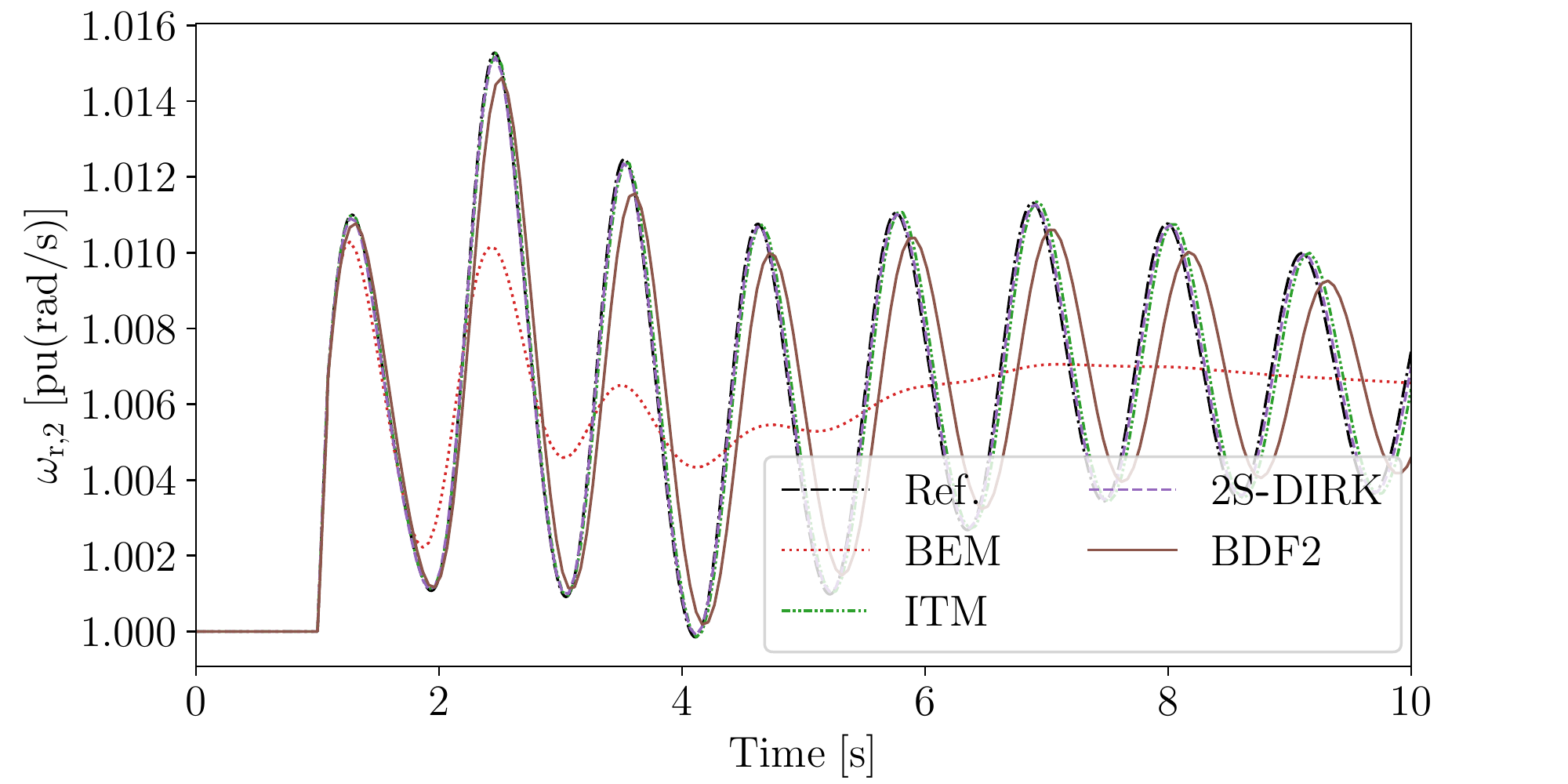}}
  \caption{WSCC system: $\omega_{{\rm r},2}$ after the fault at bus~5, $h=0.05$~s.}
  \label{fig:wscc:tds:p05}
\end{figure}

\begin{figure}[ht!]
  \centering
  \resizebox{\linewidth}{!}{\includegraphics{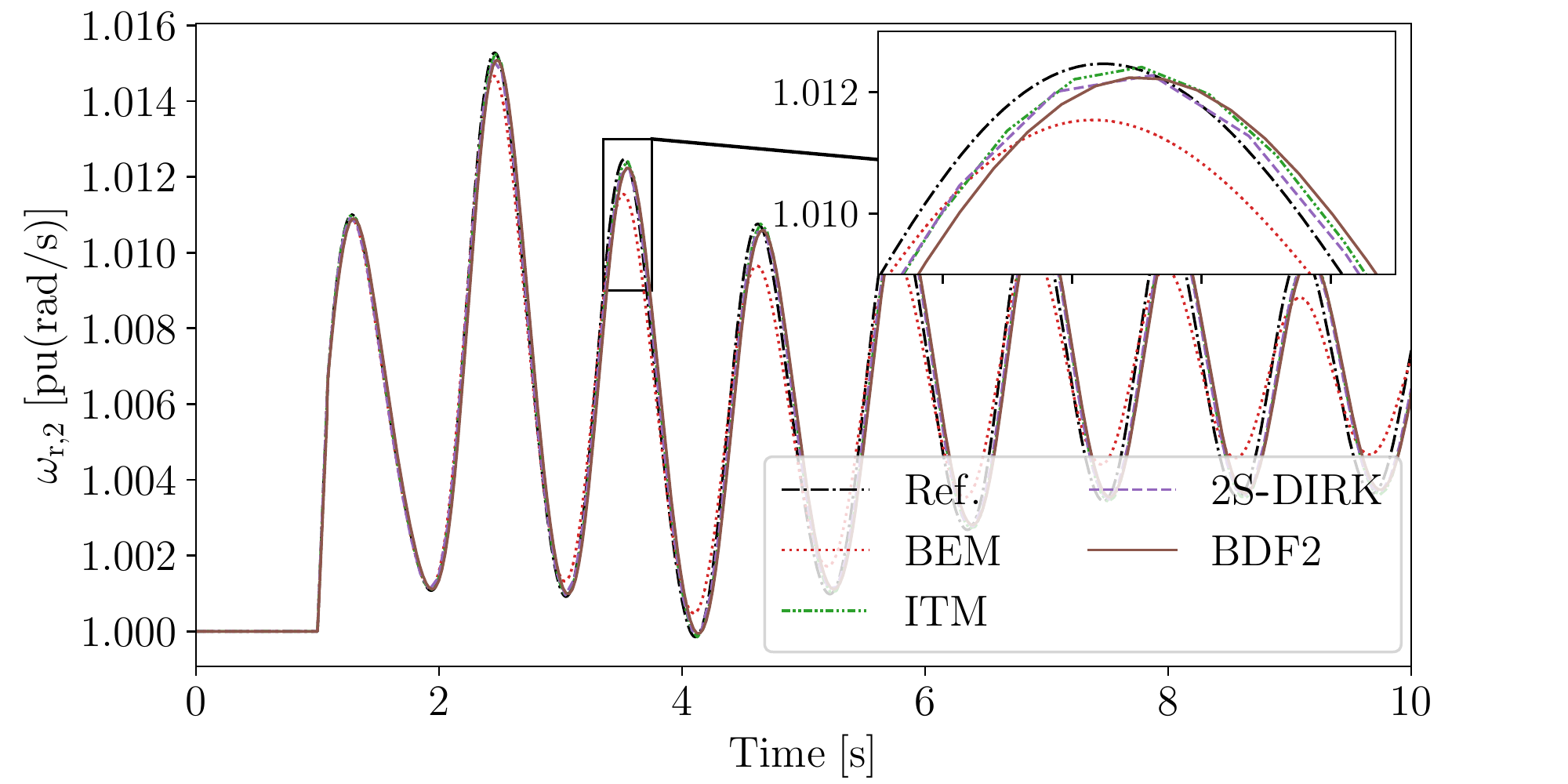}}
  \caption{WSCC system: $\omega_{{\rm r},2}$ after the fault at bus~5,
    $|d_s|=0.1$.}
  \label{fig:wscc:tds:ep1}
\end{figure}

In the remainder of this section, we focus exclusively on implicit methods, i.e.~further simulation results are provided for the
\ac{bem}, \ac{itm}, \ac{2sdirk} and \ac{bdf2}.
We carry out a non-linear time domain simulation considering a three-phase short-circuit at bus~5. The fault occurs at $t=1$~s and is
cleared after $80$~ms by tripping the line that connects buses 5 and 7. The system is integrated using $h=0.05$~s.  The response of the
rotor speed of the \ac{sg} at bus~2 ($\omega_{{\rm r},2}$), i.e., of the variable mostly participating to the dominant system mode, is shown in Fig.~\ref{fig:wscc:tds:p05}.  For comparison, we have
included a reference trajectory which represents an accurate integration of the system.\footnote{The reference trajectory in \ac{tdi} results of this paper are obtained using the \ac{2sdirk} with $h=0.001$~s.}  The trajectories in Fig.~\ref{fig:wscc:tds:p05} are consistent with the results of Table~\ref{tab:wscc:zetah},
confirming that the small-disturbance analysis results provide a rough
yet accurate estimation of the damping distortion introduced by
\ac{tdi} methods during the simulation.  Considering the same
disturbance, we simulate the system with the values of $h$ that
correspond to $|d_s|=0.1$ of the dominant mode, as obtained in
Table~\ref{tab:wscc:zetah}.  The response of the rotor speed of the
\ac{sg} at bus~2 in this case is shown in Fig.~\ref{fig:wscc:tds:ep1}.
As expected, integrating the system under a certain magnitude of
numerical distortion leads to similar trajectories for all methods.
Yet, the trajectories present some differences, since the distortion
of each method is not in the same direction with the others (see
e.g. Fig.~\ref{fig:wscc:locus12}).  For example, the oscillation
obtained with the \ac{bem} appears to be the most suppressed since the
direction of its distortion introduces the largest overdamping.

\subsection{All-Island Irish Transmission System}
\label{sec:aiits}

This section provides simulation results on a $1,479$-bus model of the
\ac{aiits}.  The topology and steady-state operation data of the
system have been provided by the Irish transmission system operator,
EirGrid Group.  Dynamic data have been determined based on current
knowledge about the technology of generators and controllers.  The
system comprises $796$~lines, $1,055$~transformers, $245$~loads,
$22$~\acp{sg}, with \acp{avr} and \acp{tg}, $6$~power system
stabilizers and $176$~wind generators.  The model has $1,443$ state
and $7,197$ algebraic variables.

\begin{table*}[ht!]
  \renewcommand{\arraystretch}{1.15}
  \centering
  \caption{ \ac{aiits}: Cumulative $\omega_{{\rm r},2}$ trajectory mismatches introduced by TDI methods, $h=0.1$~s.}
\label{tab:multiple}
\begin{threeparttable}
\begin{tabular}{ll|llll}
\toprule 
{Oper. point} & {Disturbance} & {BEM} & {ITM}
& {2S-DIRK} & {BDF2} \\
\midrule 
Base case & \ac{sg} outage  
& $1.8 \cdot 10^{-2}$ & $4.3 \cdot 10^{-4}$ 
& $2.0 \cdot 10^{-4}$ & $8.8 \cdot 10^{-3}$ \\
& Load trip 
& $6.4 \cdot 10^{-3}$ & $2.2 \cdot 10^{-3}$ 
& $1.2 \cdot 10^{-4}$ & $2.2 \cdot 10^{-3}$ \\
& 3-phase fault 
& $8.7\cdot 10^{-3}$ & $2.7 \cdot 10^{-3}$
& $1.3\cdot 10^{-3}$ & $9.0 \cdot 10^{-3}$ \\
& EWIC trip
& $2.5\cdot 10^{-2}$ & $1.1\cdot 10^{-3}$
& $5.4\cdot 10^{-4}$ & $1.2\cdot 10^{-2}$ \\
\midrule 
$+5$\% load & \ac{sg} outage 
& $5.4\cdot 10^{-1}$ & $5.6\cdot 10^{-2}$
& $5.5\cdot 10^{-2}$ & $7.3\cdot 10^{-2}$ \\
& Load trip 
& $1.0\cdot 10^{-1}$ & $1.7\cdot 10^{-2}$
& $8.2\cdot 10^{-3}$ & $4.3\cdot 10^{-2}$ \\
& 3-phase fault 
& $3.2\cdot 10^{-1}$ & $4.4\cdot 10^{-2}$
& $3.3\cdot 10^{-2}$ & $1.8\cdot 10^{-1}$ \\
& EWIC trip 
& $4.6\cdot 10^{-2}$ & $1.2\cdot 10^{-2}$
& $6.0\cdot 10^{-3}$ & $3.9\cdot 10^{-2}$ \\
\midrule 
$-5$\% load & \ac{sg} outage 
& $1.8\cdot 10^{-2}$ & $6.7\cdot 10^{-4}$
& $3.1\cdot 10^{-4}$ & $9.0\cdot 10^{-3}$ \\
& Load trip 
& $5.5\cdot 10^{-3}$ & $2.0\cdot 10^{-3}$
& $1.1\cdot 10^{-4}$ & $2.0\cdot 10^{-3}$ \\
& 3-phase fault 
& $4.2\cdot 10^{-2}$ & $1.3\cdot 10^{-2}$
& $6.4\cdot 10^{-3}$ & $4.3\cdot 10^{-2}$ \\
& EWIC trip 
& $1.2\cdot 10^{-1}$ & $5.7\cdot 10^{-3}$
& $3.7\cdot 10^{-3}$ & $5.9\cdot 10^{-2}$ \\
\bottomrule 
\end{tabular}
\end{threeparttable}\end{table*}

\begin{figure}[ht!]
  \centering
  \resizebox{\linewidth}{!}{\includegraphics{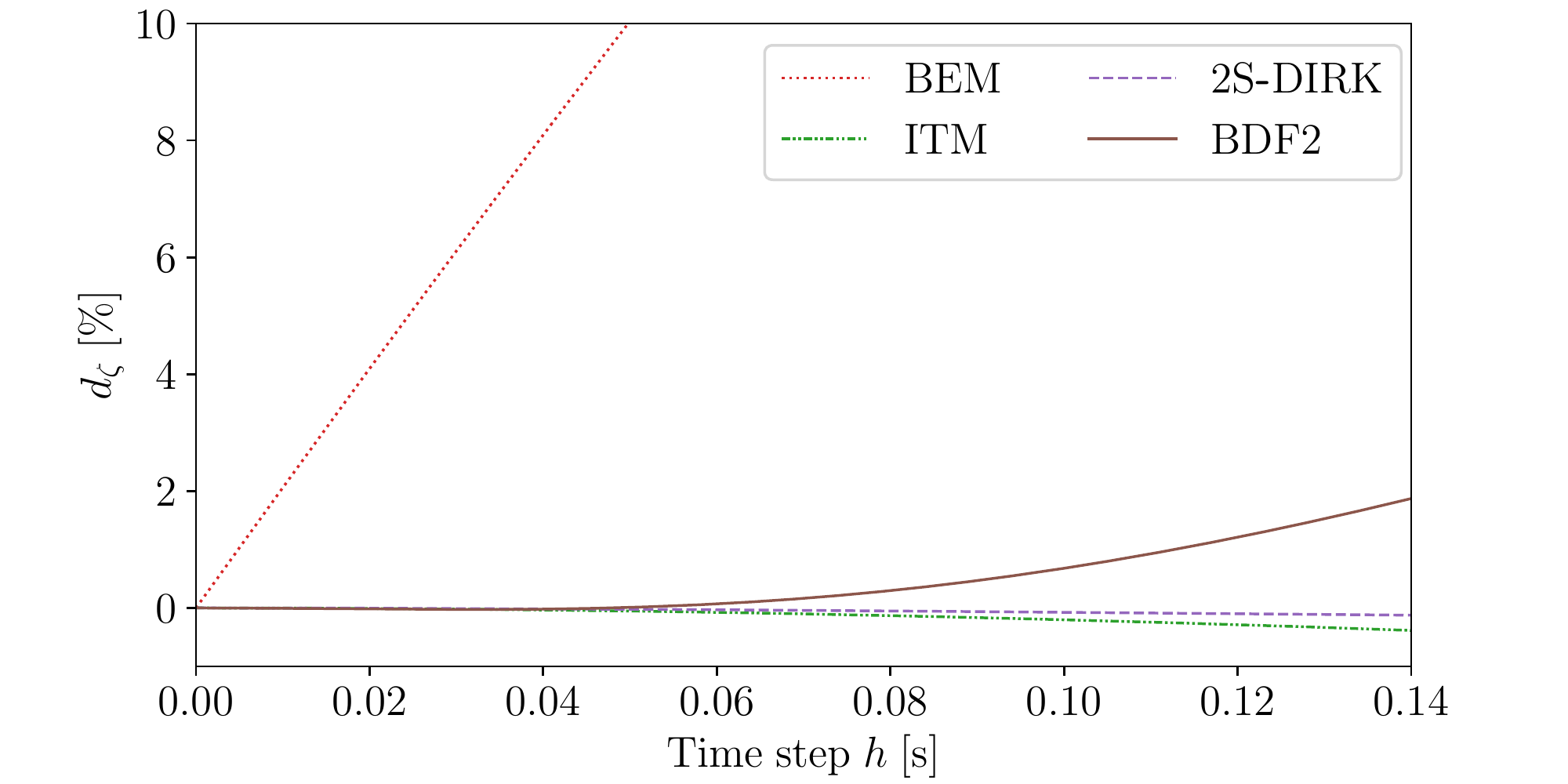}}
  \caption{\ac{aiits}: Damping distortion $d_\zeta$, MCEM.}
  \label{fig:aiits:mode1:dzeta}
\end{figure}

Eigenvalue analysis shows that the system is stable around the
examined equilibrium point.  The fastest and slowest dynamic modes
have exponential decay rates $-99,900.1$ and $-0.077$, respectively,
and thus the stiffness ratio of the model is
$\mathcal{S} = 1.3 \cdot 10^6$.
We consider the most poorly damped electromechanical mode of the system, i.e., the local oscillation of the \ac{sg} connected to bus~$507$.  In the eigenvalue analysis results, this mode is represented by the complex pair $-0.3042\pm \jj4.1426$ with damping
ratio $7.32$\%.  Hereafter, we will refer to this mode as MCEM (Most Critical Electromechanical Mode).
The magnitude of numerical distortion of the damping of the MCEM as a function of $h$ for the \ac{bem}, \ac{itm}, \ac{2sdirk} and \ac{bdf2} is shown in Fig.~\ref{fig:aiits:mode1:dzeta}.

\begin{table}[ht!]
  \centering
  \renewcommand{\arraystretch}{1.05}
  \caption{\ac{aiits}: Time step for $|d_s|=0.1$, MCEM.}
  \label{tab:aiits:mode1:h}
  \begin{threeparttable}
    \begin{tabular}{lccccccc}
      \toprule[0.1pt]
      &\ac{bem} & \ac{itm} & \ac{2sdirk}& \ac{bdf2} \\
      \midrule[0.1pt]
      $h$~[s] & 0.011 & 0.131 & 0.189  & 0.066  \\
      \bottomrule[0.1pt]%\bottomrule[0.1pt]
    \end{tabular}
  \end{threeparttable}
  \vspace{-3mm}
\end{table}

\begin{figure}[ht!]
  \centering
  \resizebox{\linewidth}{!}{\includegraphics{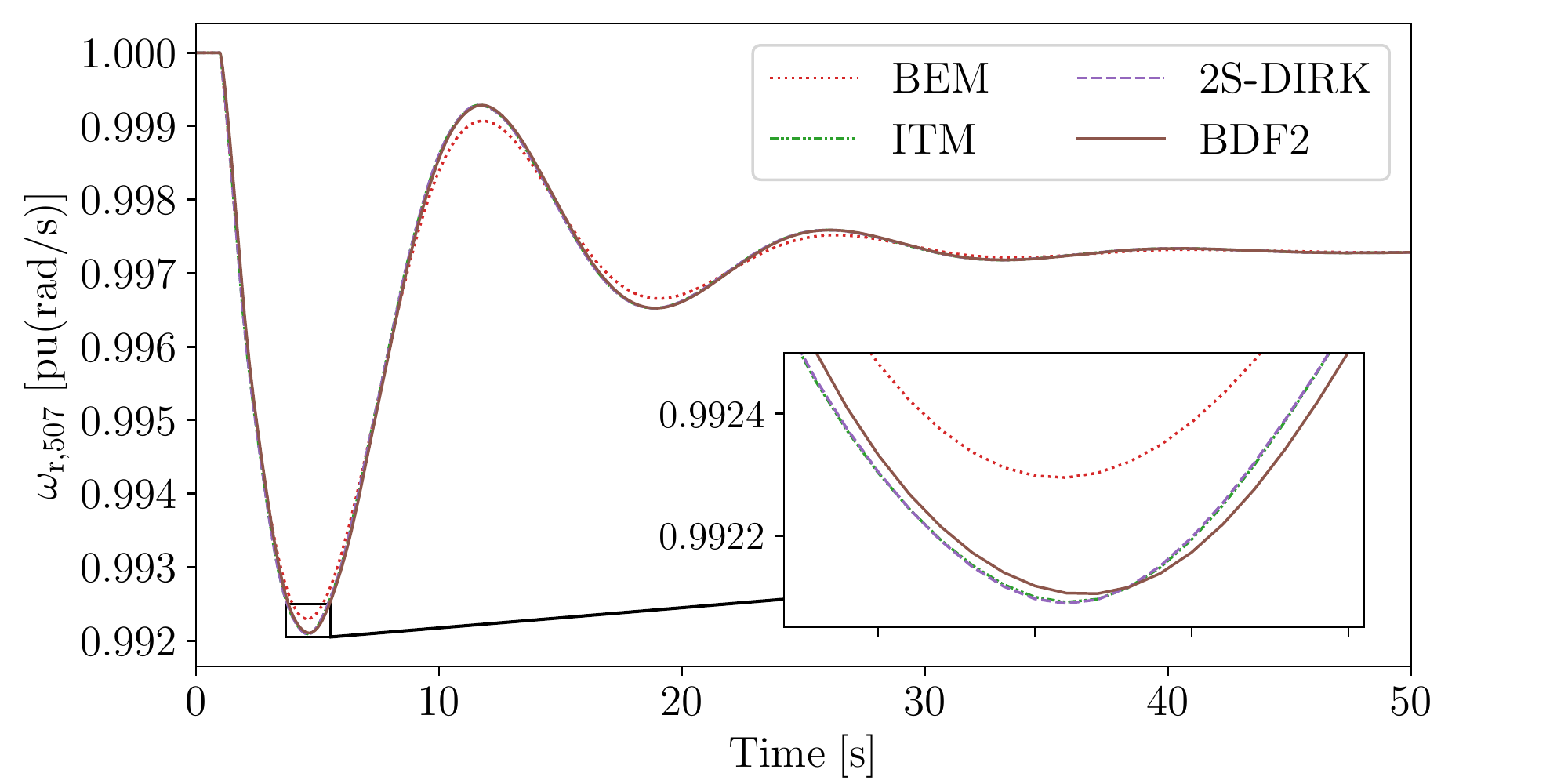}}
  \caption{\ac{aiits}: $\omega_{{\rm r},2}$ after \ac{sg} outage, $|d_s|=0.1$ (MCEM).}
  \label{fig:aiits:mode1:tds1}
\end{figure}

The time steps that correspond to $|d_s|=0.1$ for the MCEM are given
in Table~\ref{tab:aiits:mode1:h}.  Using these time step values we provide a comparison of the four implicit numerical methods by executing a non-linear \ac{tdi}, assuming the loss of the \ac{sg}
connected to bus~684 at $t=1$~s.  The response of the rotor speed of
the \ac{sg} at bus~507 following the disturbance is shown in
Fig.~\ref{fig:aiits:mode1:tds1}.  As expected, all methods provide a similar response and a small deviation from the reference trajectory.
The trajectory obtained with the \ac{bem} appears to be more damped
than the others, which was also to be expected (see
Section~\ref{sec:wscc}).

\begin{figure}[ht!]
  \centering
  \resizebox{\linewidth}{!}{\includegraphics{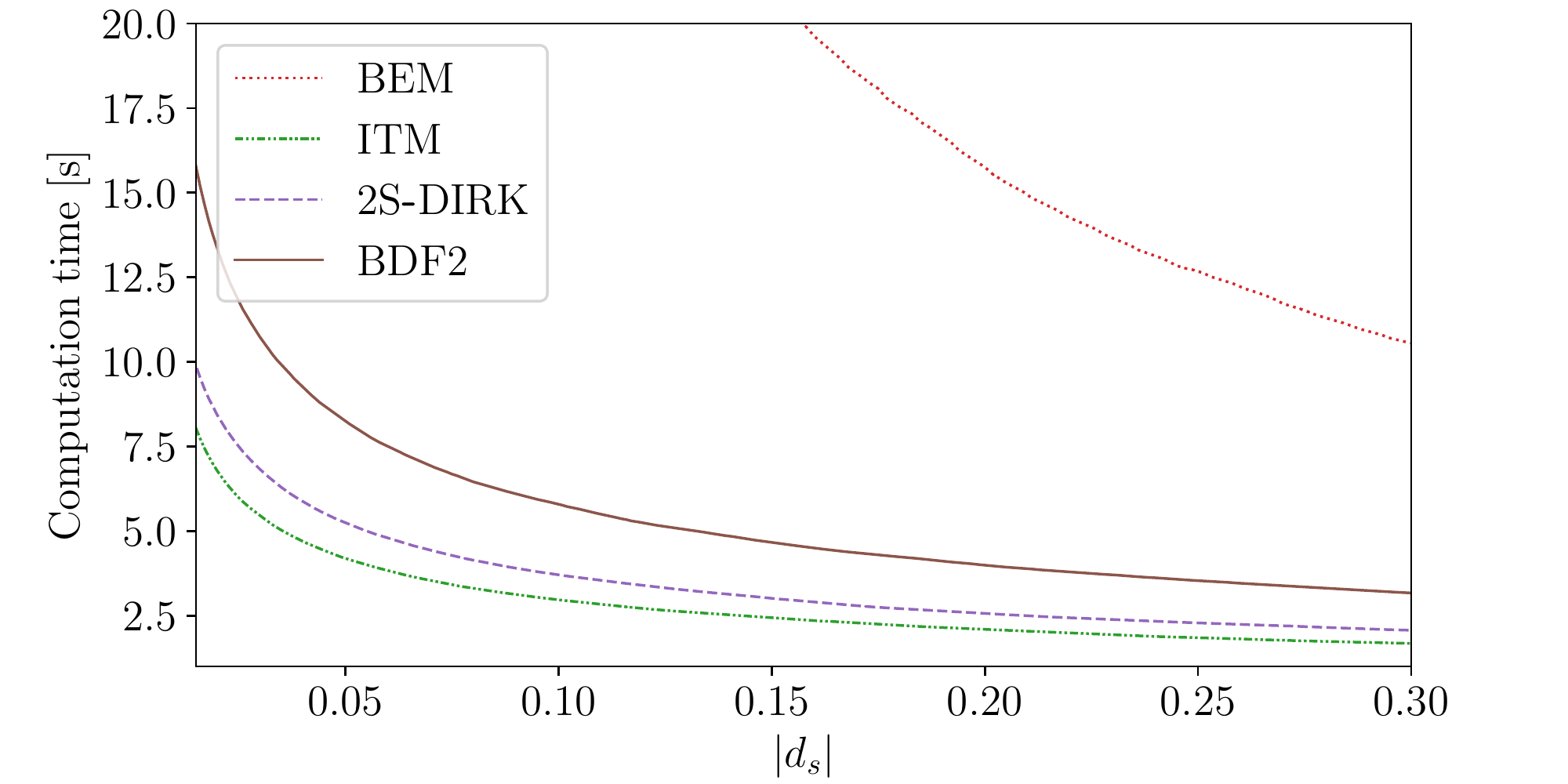}}
  \caption{\ac{aiits}: $|d_s|$ of MCEM vs \ac{tdi} total computation time.}
  \label{fig:aiits:mode1:times}
\end{figure}

A relevant remark is that evaluating different methods under the same
amount of numerical distortion can be employed as a means for their
fair computational comparison.  As an example, for each implicit
\ac{tdi} method we vary $|d_s|$ for the MCEM and for every value we calculate the corresponding time step $h$.  Using this step value we
integrate the system and compute the computational time required to complete the simulation.

The results shown in Fig.~\ref{fig:aiits:mode1:tds1} indicate that, for the examined scenario, the \ac{itm} is the method that takes the lowest total computational time.  Yet, as $|d_s|$ increases, the relative difference of the \ac{itm} with respect to the other methods decreases.  For large time steps, in fact, the \ac{itm} shows large sustained numerical oscillations which in turn lead to an increased number of required iterations per step, while the opposite is true for the methods that introduce overdamping, i.e.,~they require less iterations.

We note that considering a single dynamic mode is assumed in Figs.~\ref{fig:wscc:error} and \ref{fig:aiits:mode1:dzeta} for the sake of
simplicity, but this is not a limitation of the proposed approach, since the analysis can be extended to take into account a group of critical modes, or even all system modes.

Finally, in addition to the 
operating condition assumed so far in this example 
(base case), we consider two more. The new operating conditions are obtained through a $137$~MW
load increase/decrease
(corresponding to $5$\% 
increase/decrease of the system's total power consumption). For each operating condition, we consider four different disturbances, 
(i) outage of $90$~MW synchronous generation at bus~684, 
(ii) loss of a total of $101.4$~MW power consumption connected to buses~1-5,
(iii) three-phase fault at bus~1238, cleared by tripping a transmission line connected to the faulted bus after $100$~ms. 
(iv) loss of the VSC-HVDC link East-West Inter-connector~(EWIC) that connects the \ac{aiits} to Great Britain's transmission system.

The system is integrated with the \ac{bem}, the \ac{itm}, the \ac{2sdirk}, and the 
\ac{bdf2}, for $15$~s and using $h=0.1$~s.
In every simulation, the disturbance is applied at $t=1$~s. Table~\ref{tab:multiple} shows, for each method and scenario, the magnitude of the error of the rotor speed $\omega_{{\rm r},2}$ with respect to the reference trajectory, cumulated for the simulation period. Moreover, the values of $|d_s|$ for the mode to which $\omega_{{\rm r},2}$ mostly participates in, i.e.~the MCEM, 
are given in Table~\ref{tab:aiits:mode1:ds}. 
The values in Table~\ref{tab:aiits:mode1:ds} are
determined considering the base case operating condition.  Results further confirm the suitability of the proposed approach in providing indicative and useful, yet not absolute measures of the numerical distortion introduced by \ac{tdi} methods. Of course, repeating the \ac{sssa} when the operating condition is varied, would allow a further improvement of the precision of the measures derived.
\begin{table}[ht!]
\centering
\renewcommand{\arraystretch}{1.05}
\caption{ \ac{aiits}: $|d_s|$ for $h=0.1$~s, MCEM.}
\label{tab:aiits:mode1:ds}
\begin{threeparttable}
\begin{tabular}{lccccccc}
 \toprule[0.1pt]
 &\ac{bem} & \ac{itm} & \ac{2sdirk}& \ac{bdf2} \\
  \midrule[0.1pt]
  $|d_s|$ & 0.810 & 0.058 & 0.029 & 0.208 \\
  \bottomrule[0.1pt]
\end{tabular}
\end{threeparttable}
\vspace{-3mm}
\end{table}

Overall, results support the validity of the proposed approach in providing indicative accuracy measures for the non-linear system model.

\section{Conclusions}
\label{sec:conclusion}

The paper provides a framework based on \ac{sssa} to study the numerical distortion introduced by explicit and implicit integration schemes when applied for the simulation of power system dynamics.  The
proposed framework is implemented using a general formulation which covers the most important families of integration methods, including \ac{rk} and linear multistep methods.  Results indicate that adopting the proposed approach, one is able to provide useful upper time step bounds to satisfy certain accuracy criteria, as well as to compare different methods in a fair way.

Future work will exploit the proposed framework to provide new insights on the accuracy of multirate methods, e.g.~see \cite{shu2017multirate,4470562,moreira2006multirate}, as well as to evaluate the ability and limitations of common integration schemes to accurately cope with time delays and stochastic processes.

\appendix
\section{Proofs of Propositions}
\label{app:proofs}

\subsection{Proof of Proposition~\ref{proposition:tdi:pencil}}

%\color{red}
We consider \eqref{eq:tdi:implicit:lin} and for simplicity we use the notation $\frac{\partial\bfg \eta}{\partial\xs_{t}}=\bfb A_0$, $\frac{\partial\bfg \eta}{\partial\xs_{t-h}} \bfb A_h$, $\frac{\partial\bfg \eta}{\partial\xs_{t-a_1h}}=\bfb A_{a_1h}$, $\frac{\partial\bfg \eta}{\partial\xs_{t-a_2h}}=\bfb A_{a_2h}$, $\ldots$, $\frac{\partial\bfg \eta}{\partial\xs_{t-a_\rho h}}=\bfb A_{a_\rho h}.$ 
Let also $\epsilon>0$, so that $a_i=c_i \epsilon$, 
$c_i \in \mathbb{N}^*$.  We set:

\begin{equation}
\begin{aligned}%\label{eq:tdi:implicit:lin:1}
\nonumber
  \bfg y_t^{[0]} &= \Delta \xs_{t}\\
   \bfg y_t^{[h]} &= \Delta \xs_{t-h}\\
 %  \bfg y_t^{[2h]} &= \Delta \xs_{t-2h} \\
   &\vdots \\
   \bfg y_t^{[a_1h]} &= \Delta \xs_{t-a_1h}\\
    \bfg y_t^{[(a_1+\epsilon)h]} &= \Delta \xs_{t-(a_1+\epsilon)h}\\
        \bfg y_t^{[(a_1+2\epsilon)h]} &= \Delta \xs_{t-(a_1+2\epsilon)h}\\
   &\vdots \\ 
   \bfg y_t^{[a_2h]} &= \Delta \xs_{t-a_2h} \\
    \bfg y_t^{[(a_2+\epsilon)h]} &= \Delta \xs_{t-(a_2+\epsilon)h}\\
   &\vdots \\ 
    \bfg y_t^{[a_{\rho-1}h]} &= \Delta \xs_{t-a_{\rho-1}h} \, .
\end{aligned}  
\end{equation}

and
\begin{equation}
\begin{aligned}
\label{eq:tdi:implicit:lin:2} \nonumber
  \bfg y_{t-h}^{[0]} &= \Delta \xs_{t-h}=\bfg y_t^{[h]}\\
   \bfg y_{t-h}^{[h]} &= \Delta \xs_{t-2h}=\bfg y_t^{[2h]}\\
   \bfg y_{t-h}^{[2h]} &= \Delta \xs_{t-3h}=\bfg y_t^{[3h]}\\
   &\vdots \\
   \bfg y_{t-h}^{[a_1h]} &= \Delta \xs_{t-(a_1+\epsilon)h}=\bfg y_t^{[(a_1+\epsilon)h]}\\
    \bfg y_{t-h}^{(a_1+\epsilon)h} &= \Delta \xs_{t-(a_1+2\epsilon)h}=\bfg y_t^{[(a_1+2\epsilon)h]}\\
   \vdots \\ 
   \bfg y_{t-h}^{[a_2h]} &= \Delta \xs_{t-(a_2+\epsilon)h}=\bfg y_t^{[(a_2+\epsilon)h]}\\
    \bfg y_{t-h}^{[(a_2+\epsilon)h]} &= \Delta \xs_{t-(a_2+2\epsilon)h}=\bfg y_t^{[(a_1+2\epsilon)h]}\\
   &\vdots \\ 
   \bfb A_{a_\rho h} \bfg y_{t-h}^{[a_{\rho-1}h]} &=\bfb A_{a_\rho h} \Delta \xs_{t-a_\rho h} \, .
\end{aligned}      
\end{equation}
The last matrix equation can be written as: 
\begin{align}
\label{eq:tdi:implicit:lin:3}
 \bfb A_{a_\rho h} \bfg y_{t-h}^{[a_{\rho-1}h]} =
   &-\bfb A_0\Delta \xs_{t}-\bfb A_h\Delta \xs_{t-h}-
   \bfb A_{a_1h}\Delta \xs_{t-a_1h}
  % -\bfb A_{a_2h}\Delta \xs_{t-a_2h}
  \nonumber \\ &
   -\ldots-\bfb A_{a_{\rho-1} h}\Delta\xs_{t-a_{\rho-1} h}
   \, .
\end{align}
or, equivalently,
\begin{align}
\label{eq:tdi:implicit:lin:4}
 \bfb A_{a_\rho h} \bfg y_{t-h}^{[a_{\rho-1}h]} =&
   -\bfb A_0\bfg y_t^{[0]}
   -\bfb A_h\bfg y_t^{[h]}-\bfb A_{a_1h}\bfg y_t^{[a_1h]}
  % -\bfb A_{a_2h}\Delta \xs_{t-a_2h}
  \nonumber \\&
   -\ldots-\bfb A_{a_{\rho-1} h}\bfg y_t^{[a_{\rho-1} h]} \, .
\end{align}    
Using the above matrix equations we arrive to system \eqref{eq:tdi:sssa} which is equivalent to \eqref{eq:tdi:implicit:lin}, where:
\begin{align}
    & \tdiE=
    \left[
    \begin{array}{cc}
    \bfb I %_{(\rho-1)(\rho-2)\nxy}
    &\bfb 0%_{(\rho-1)(\rho-2)\nxy,\nxy}
    \\
    \bfb 0 %_{\nxy,(\rho-1)(\rho-2)\nxy}
    &\bfb A_{a_\rho h}
    \end{array}
    \right]\,, \\
    &\tdiA=
    \left[
    \begin{array}{cc}
    \bfb 0%_{(\rho-1)(\rho-2)\nxy,\nxy}
    &\bfb I%_{(\rho-1)(\rho-2)r}
    \\
    -\bfb A_0&\bfb A^\dagger%_{a_\rho h}
    \end{array}
    \right] \, ,
\end{align}
and:
\begin{equation}
    \bfb A^\dagger=
    \left[
    \begin{array}{cccccccc}
    -\bfb A_h& 
    \bfb 0&
    \dots
   -\bfb A_{a_1h}&
   \bfb 0&
   \dots&
  % -\bfb A_{a_2h}\Delta \xs_{t-a_2h}
   -\bfb A_{a_{\rho-1} h}&
   \bfb 0
    \end{array}
    \right] \, ,
    \nonumber
\end{equation}
where $\bfb 0$, $\bfb I$ are the zero and identity matrix with proper dimensions. %=\bfb 0_{(\rho-1)(\rho-2)\nxy,\nxy}$ and: 
\begin{equation}
\begin{aligned}
\bfb y_t\T =& [(\bfg y_t^{[0]})\T  \ \ \ 
(\bfg y_t^{[h]})\T  \ \ \ 
(\bfg y_t^{[2h]})\T  \ \ \dots  \ \ 
(\bfg y_t^{[a_1h]})\T  \\ & 
(\bfg y_t^{[(a_1+\epsilon)h]})\T  \ \ \dots  \ \
(\bfg y_t^{[a_2h]})\T  \ \ \ 
(\bfg y_t^{[(a_2+\epsilon)h]})\T  \\ 
& \ \ \dots \ \
(\bfg y_t^{[a_{\rho-1}h]})\T ] \, . \\
\end{aligned}
\end{equation}
Then, we have that the pencils
$s^{a_\rho h}\bfb A_{a_\rho h}+\ldots+s^{a_2h}\bfb A_{a_2h}+s^{a_1h}\bfb A_{a_1h}+s^h\bfb A_h+\bfb A_0$, $s\tdiE-\tdiA$ of systems \eqref{eq:tdi:implicit:lin}, \eqref{eq:tdi:sssa} respectively, have exactly the same finite eigenvalues, see 
\cite{book:eigenvalue}. The proof is completed.

\subsection{Propositions~\ref{proposition:moebius} and \ref{proposition:itm}}

{\proposition
  {
    Recall the general form of the M{\"o}bius transformation \cite{moebius}:
    \begin{equation}
      \label{eq:moebius1}
      s = \frac{az+b}{cz+d} \, ,
      \quad a, 
      b,c,d\in\mathbb{C},\quad ad-bc\neq 0
      \, .
    \end{equation}
    If \eqref{eq:moebius1} is applied to the matrix pencil $s \bfb E - \bfb A$, it
    leads to the family of pencils:
    \begin{equation}
      \label{eq:mpencils}
      z(a\bfb E-c\bfb A)-(d\bfb A-b\bfb E)
      \, .
    \end{equation}
  }
  \label{proposition:moebius}
  %	\vspace{0.5mm}
}
      %       The proof of Proposition~\ref{proposition:moebius}
      %       is provided in Appendix~\ref{app:proofs}.
\proof	
% Consider the general form of the M{\"o}bius transformation given by \eqref{eq:moebius1}. 
First, the restriction 
$ad-bc\neq 0$ in \eqref{eq:moebius1} is 
necessary since, if $ad=bc$ then $s$ is constant, which is not possible. 
Then, recall that the eigenvalues of the pencil of \eqref{eq:dae:lin} are the solutions of \eqref{eq:dae:chareq},
% by the roots of
% the characteristic polynomial
% 
% \[
% ${\rm det}
% (s\bfb E-\bfb A\big)$, 
% \, ,
% \]
% 
whereby applying 
\eqref{eq:moebius1} we get:
\[
  {\rm det}
  \Big( \frac{az+b}{cz+d}\bfb E-\bfb A\Big )=0 \, ,
\]
or, equivalently, by using determinant properties:
% \[
% {\rm det}
% \big((az+b)\bfb E-(cz+d)\bfb A\big)=0 \, ,
% \]
% or, equivalently: 
\[
  {\rm det}
  ((a\bfb E-c\bfb A)z-(d\bfb A-b\bfb E))=0 \, ,
\]
i.e.~the characteristic equation of a linear system with pencil:
% \[
% (a\bfb E-c\bfb A)\hat{\xs}'(t)=(d\bfb A-b\bfb E)\hat{\xs}(t) \, ,
% \]
% with pencil 
% 
% 
\begin{equation}
  \label{eq:mpencils2}
  z(a\bfb E-c\bfb A)-(d\bfb A-b\bfb E)
  \, .
\end{equation}
% 
% where
% \[
% \tilde{\bfb E}=a\bfb E-c\bfb A, \quad %\tilde{A}=d\bfb A-b\bfb E.
% \]

The proof is completed. Then, we may obtain as
special cases the matrix pencils of
the \ac{fem}, for
$a=1$, $b=-1$, $c=0$, $d=h$;
the \ac{bem}, for
$a=1$, $b=-1$, $c=h$, $d=0$;
the \ac{itm}, for 
$a=1$, $b=-1$, $c=0.5h$, $d=0.5h$.
% which corresponds to the transform
% $s=\frac{2}{h}\frac{z-1}{z+1}$.
% i.e.~the bilinear transform.
% 

Note that the family of pencils \eqref{eq:mpencils}
corresponds to a family of linear 
discrete-time systems 
in the form:
\begin{equation}
  \label{eq:map_o1:sssa2}
  (a \bfb E - c \bfb A ) \Delta\xs_{t} =   
  (d \bfb E - b \bfb A) \Delta\xs_{t-h}
  \, .
\end{equation}

\vspace{1mm}
{\proposition
  {
    Consider system \eqref{eq:dae:lin}
    with $ad-bc\neq 0$.
    % and the discrete map \eqref{eq:map_o1:sssa2}.  
    If one of the following conditions holds:
    \begin{equation}
      \label{holds}
      a = b \ \text{and}
      \  
      c =  - d \, , 
      \quad
      \text{or} 
      \quad
      a = -b \  \text{and}
      \ 
      c = d \, ,
    \end{equation}
    then for a stable equilibrium state of \eqref{eq:dae:lin}, the 
    magnitude of the spectral radius of the matrix pencil of each discrete-time system in the form of \eqref{eq:map_o1:sssa2} is $<1$.
  }
  \label{proposition:itm}
  \vspace{1mm}
}

\proof	

For a stable equilibrium state of \eqref{eq:dae:lin}, we have 
${\rm Re}(s)<0$, for every finite eigenvalue $s\in\mathbb{C}$ and hence $s+\bar{s}<0$,
where $\bar{s}$ is the complex conjugate of $s$.
Substituting $s={(az+b)}/({cz+d})$ we get: 
\[
  \frac{az+b}{cz+d}+\frac{a\bar{z}+b}{c\bar{z}+d}<0 \, ,
\]
or, equivalently:
\[
  (c\bar{z}+d)(az+b)+(a\bar{z}+b)(cz+d)<0 \, ,
\]
or, equivalently, by taking into account that $\bar{z}z=|z|^2$:
\[
  2ac|z|^2+2bd+(ad+bc)(\bar{z}+z)<0 \, ,
\]
%\color{blue}
This means that the set $\{{\rm Re}(s)<0,\ \forall s\in\mathbb{C}\}$ maps to the set $\{ac|z|^2+bd+(ad+bc){\rm Re}(z)<0,
\ \forall z\in\mathbb{C}\}$.
If we apply conditions \eqref{holds}, we have $ac+bd=0$ and $ad+bc=0$ which is equal to ${bd}/{ac}=-1$. 
Hence, the above relation takes the form:
\[
2ac|z|^2<-2bd \, ,
\]
and consequently $|z| < 1$.
Thus, through \eqref{eq:dae:lin} and under the conditions \eqref{holds}, the set $\{{\rm Re}(s)<0,\ \forall s\in\mathbb{C}\}$ maps to the set $\{|z|<1,
\ \forall z\in\mathbb{C}\}$ %through \eqref{eq:dae:lin} when \eqref{holds} holds 
and consequently the stability of this continuous time system can be studied through the stability of the discrete-time system \eqref{eq:map_o1:sssa2}.
\color{black}
Hence for a stable equilibrium state of \eqref{eq:dae:lin}, we obtain that the magnitude of the spectral radius of the pencil of each discrete-time system in
the form of \eqref{eq:map_o1:sssa2} is $<1$. 
The proof is completed.

\bibliographystyle{IEEEtran}
\bibliography{references}

\newpage
  
  \begin{biography}
    [{\includegraphics[width=1in,height=1.25in,clip,keepaspectratio]{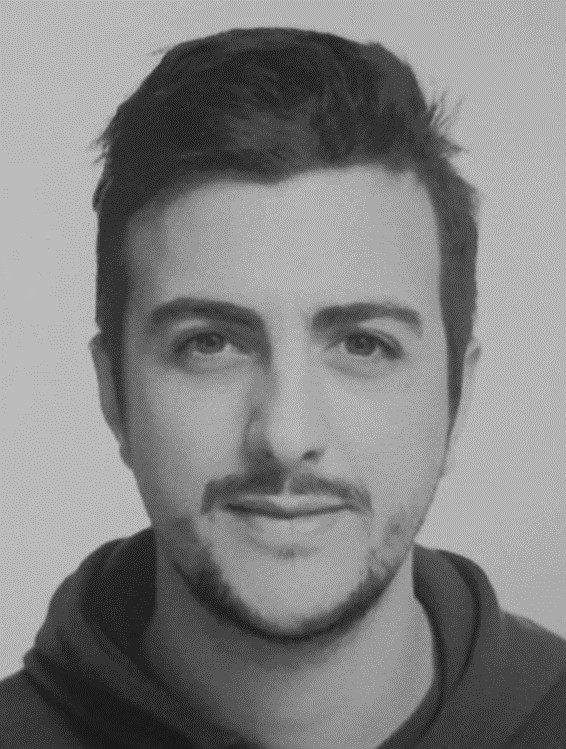}}]
    {Georgios Tzounas} (M'21) received the Diploma (M.E.)~degree in Electrical and
    Computer Engineering from the National Technical Univ. of Athens, Greece, in 2017, and the Ph.D. degree in Electrical Engineering from
    Univ. College Dublin (UCD), Ireland, in 2021. 
    From Jan. to Apr.~2020, he
    was a Visiting Researcher at Northeastern Univ., Boston, MA.  
    He is currently a Senior Power Systems Researcher at UCD, working on the EU H2020 project “EdgeFLEX”.  His research
    interests include modelling, stability analysis and control of power systems.
  \end{biography}
  
  \begin{biography}
    [{\includegraphics[width=1in,height=1.25in,clip,keepaspectratio]{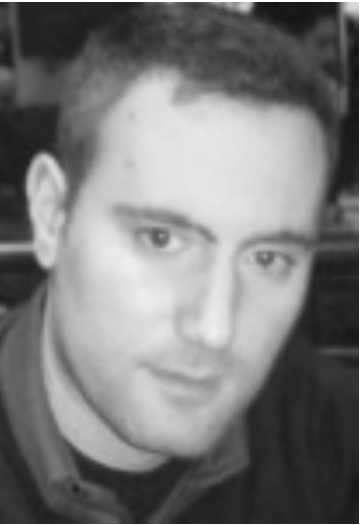}}]
    {Ioannis Dassios} received his Ph.D. in Applied Mathematics from the
    Dpt of Mathematics, Univ. of Athens, Greece, in 2013.  He worked as
    a Postdoctoral Research and Teaching Fellow in Optimization at the
    School of Mathematics, Univ.~of Edinburgh, UK.  He also worked as a
    Research Associate at the Modelling and Simulation Centre,
    University of Manchester, UK, and as a Research Fellow at MACSI,
    Univ.~of Limerick, Ireland. He is currently a UCD Research Fellow at UCD, Ireland.
  \end{biography}
  
  \begin{biography}
    [{\includegraphics[width=1in,height=1.25in,clip,keepaspectratio]{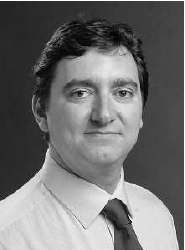}}]
    {Federico Milano} (F'16) received from the University of Genoa,
    Italy, the ME and Ph.D.~in Electrical Engineering in 1999 and 2003,
    respectively.  From 2001 to 2002, he was with the Univ.~of Waterloo,
    Canada.  From 2003 to 2013, he was with the Univ.~of Castilla-La
    Mancha, Spain.  In 2013, he joined the Univ.~College Dublin,
    Ireland, where he is currently Professor of Power Systems Control
    and Protections and Head of Electrical Engineering.  His research
    interests include power systems modeling, control and stability analysis.
  \end{biography}

\vfill

\end{document}